\documentclass[12pt]{article}
\usepackage{amsfonts}
\usepackage{latexsym}
\usepackage{amsmath}
\usepackage{amssymb}
\usepackage{amssymb}
\catcode`\@=11
\newcount\hour
\newcount\minute
\newtoks\amorpm \hour=\time\divide\hour by 60\minute
=\time{\multiply\hour by 60 \global\advance\minute by-\hour}
\edef\standardtime{{\ifnum\hour<12 \global\amorpm={am}%
        \else\global\amorpm={pm}\advance\hour by-12 \fi
        \ifnum\hour=0 \hour=12 \fi
        \number\hour:\ifnum\minute<10
        0\fi\number\minute\the\amorpm}}
\edef\militarytime{\number\hour:\ifnum\minute<10
0\fi\number\minute}
\def\draftlabel#1{{\@bsphack\if@filesw {\let\thepage\relax
   \xdef\@gtempa{\write\@auxout{\string
      \newlabel{#1}{{\@currentlabel}{\thepage}}}}}\@gtempa
   \if@nobreak \ifvmode\nobreak\fi\fi\fi\@esphack}
        \gdef\@eqnlabel{#1}}
\def\@eqnlabel{}
\def\@vacuum{}
\def\marginnote#1{}
\def\draftmarginnote#1{\marginpar{\raggedright\scriptsize\tt#1}}
\def\draft{
        \pagestyle{plain}
        \overfullrule=2pt
        \oddsidemargin -.1truein
        \def\@oddhead{\sl \phantom{\today\quad\militarytime} \hfil
        \smash{\Large\sl DRAFT} \hfil \today\quad\militarytime}
        \let\@evenhead\@oddhead
        \let\label=\draftlabel
        \let\marginnote=\draftmarginnote
        \def\ps@empty{\let\@mkboth\@gobbletwo
        \def\@oddfoot{\hfil \smash{\Large\sl DRAFT} \hfil}
        \let\@evenfoot\@oddhead}
        \def\@eqnnum{(\theequation)\rlap{\kern\marginparsep\tt\@eqnlabel}%
        \global\let\@eqnlabel\@vacuum}  }

\renewcommand{\theequation}{\thesection.\arabic{equation}}
\renewcommand{\thefootnote}{\fnsymbol{footnote}}
\newcommand{\newsection}{    
\setcounter{equation}{0}\section}
\def\appendix#1{\addtocounter{section}{1}\setcounter{equation}{0}
\renewcommand{\thesection}{\Alph{section}}
\section*{Appendix \thesection\protect\indent \parbox[t]{11.15cm}{#1}}
\addcontentsline{toc}{section}{Appendix \thesection\ \ \ #1}}

\def \bi{\bibitem}
\def \la {\label}

\def\be{\begin{equation}}
\def\ee{\end{equation}}

\def\inphi{(\partial_a\Phi)}
\def\apphi{ |\partial_a\Phi|}

\catcode`\@=11

\def\bea{\begin{eqnarray}}
\def\eea{\end{eqnarray}}
\def\beann{\begin{eqnarray*}}
\def\eeann{\end{eqnarray*}}
\def\beq{\begin{equation}}
\def\eeq{\end{equation}}
\def\ba{\begin{array}}
\def\ea{\end{array}}
\def\ben{\begin{enumerate}}
\def\een{\end{enumerate}}

 \def \la {\label}
 \def\be{\begin{equation}}
\def\ee{\end{equation}}

\def \la {\label}

\font\mybb=msbm10 at 11pt

\def\bb#1{\hbox{\mybb#1}}

\def\bR {\bb{R}}

\def\bC {\bb{C}}

\def\e  {\epsilon}

\def \ee {\epsilon}

\def \bi{\bibitem}

\def\mt{\eta}

\def\be{\begin{equation}}
\def\ee{\end{equation}}

\def \bi {\bibitem}
\def \la{\label}

\begin{document}
\date{November 2002}
\begin{titlepage}
\begin{center}
\vspace{5.0cm}

\vspace{3.0cm} {\Large \bf Heterotic supersymmetric backgrounds with compact holonomy revisited}
\\
[.2cm]


{}\vspace{2.0cm}
 {\large
 G.~Papadopoulos
 }



{}

\vspace{1.0cm}
Department of Mathematics\\
King's College London\\
Strand\\
London WC2R 2LS, UK\\


\end{center}
{}
\vskip 3.0 cm
\begin{abstract}
We simplify the classification of supersymmetric solutions with compact holonomy of the Killing spinor equations
 of heterotic supergravity
using the field equations and the additional assumption that the 3-form flux is closed.  We determine all the fractions
of supersymmetry that the solutions  preserve and find that there is a restriction
on the number of supersymmetries  which depends on the isometry group
of the background. We  examine the geometry of spacetime in all cases. We find that the supersymmetric solutions
of heterotic supergravity
are associated with a large number of geometric structures which include 7-dimensional manifolds with $G_2$ structure,
 6-dimensional complex and almost complex
manifolds, and 4-dimensional hyper-K\"ahler, K\"ahler and anti-self-dual Weyl manifolds.

\end{abstract}
\end{titlepage}
\newpage
\setcounter{page}{1}
\renewcommand{\thefootnote}{\arabic{footnote}}
\setcounter{footnote}{0}

\setcounter{tocdepth}{1}
\tableofcontents

\setcounter{section}{0}
\setcounter{subsection}{0}
\newpage

\newsection{Introduction}

The Killing spinor equations of heterotic supergravity have been solved in all cases \cite{het1, het2}
and it has been found that
there are 61 type of solutions up to gauge transformations  tabulated in table 2 of \cite{het3}.
The solutions can be separated into two large classes depending on whether the holonomy, ${\rm hol}(\hat\nabla)$,
 of the connection $\hat\nabla$
with torsion the 3-form flux $H$ is compact or non-compact. The holonomy group describes completely
the solution of the gravitino Killing spinor equation (KSE). Each class is further subdivided. This
is because not all solutions of the gravitino KSE are also solutions of the dilatino one. If $L$ is the number
of $\hat\nabla$-parallel spinors, ie the solutions of the gravitino KSE, then typically the
number $N$ of Killing spinors,
ie solutions of both gravitino and dilatino KSEs, is $N\leq L$. Backgrounds with $N<L$ have been  called descendants in
\cite{het2}.

In the non-compact holonomy case, the solutions of the KSEs are characterized by the pair of numbers $(L,N)$.
In particular, $L=1,2,3,4,5,6,8$, and each $L$ is associated to a unique non-compact holonomy group. Moreover
 $N$ takes all values $N\leq L$ for each $L$. In addition
for every pair $(L,N)$ there is a unique type of spacetime geometry that can occur. Furthermore, it has been shown
that the geometry
of the $(L,N)$, $N\not=7$, backgrounds is a special case of that of $(N,N)$ backgrounds. It suffices therefore
to consider only those backgrounds for which all parallel spinors are Killing. This is apart from the
$(8,7)$ case which is treated separately, see also table 4. Since there is a concise description of
the geometry of $(L,L)$
backgrounds, the understanding of the geometric conditions imposed by supersymmetry is complete
for all backgrounds with non-compact holonomy.

In the compact holonomy case, the solutions of the KSEs can again be labeled by the pair  $(L,N)$, where
$L=2,4,8,16$ and $N\leq L$. But, unlike
the non-compact case, they
are not uniquely characterized by the pair $(L,N)$. In particular for a given pair $(L,N)$, there are different
types of geometry that can occur. Moreover, there is no straightforward relation between the  geometry of $(L,N)$
and $(N,N)$ backgrounds. Of course all conditions on the spacetime
geometry  that arise from the
KSEs are known  \cite{het2}.  However, they are stated in a  non-covariant manner
as an artifact of the gauging fixing process used in the context of
  spinorial geometry method \cite{ggp} to solving the dilatino KSEs. Furthermore, it is known that the spacetime
  of such supersymmetric backgrounds
  admits a Lorentzian Lie algebra action generating Killing vector fields which are nowhere vanishing. Nevertheless,
  the classification of Lorentzian Lie algebras \cite{medina, josek} has not been incorporated in the understanding
  of geometry of supersymmetric
  backgrounds.

  In this paper, we shall re-examine the dilatino Killing spinor equation in a way that it is manifestly covariant.
  However, this cannot be achieved without some additional assumptions which we shall explain later.
Nevertheless the covariant approach to solving the dilatino KSE has some advantages.
One is that we illuminate the large degeneracy of types of geometry which occurs for each pair
$(L,N)$, and we find some restriction on $N$ for a given $L$.
The other advantage is that the classification of Lorentzian Lie algebras is now naturally incorporated
in the classification of supersymmetric backgrounds.

Different sets of assumptions can be used to solve the dilatino KSE. However
 the most economical
assumption is to take $dH=0$, ie impose the Bianchi identity of the 3-form field strength, and
also use the field equations that arise as the conditions for conformal invariance at 1-loop
in the sigma model perturbation, ie the heterotic supergravity field equations\footnote{Alternatively,
we can assume the consequences of imposing $dH=0$ and the field equations but allow for $H$ not to be closed.}. The condition $dH=0$
is always valid in the heterotic case at the zeroth order of the $\alpha'$ expansion, and to all orders
if the gauge connection
is embedded in the spin connection. Moreover, it is valid to all orders for the type II common sector backgrounds.

The above assumptions have far reaching consequences. One reason for this is that one can use holonomy reduction
arguments. These are based on the observation that the descendant backgrounds, $N<L$, have more
$\hat\nabla$-parallel tensors
than those with $N=L$ \cite{het2}. As a result, the holonomy of $\hat\nabla$ for the descendant backgrounds
  reduces. Examining the pattern of holonomy reduction,
one can determine some components of $H$ which in turn allow for the solution of the dilatino KSE.
Another consequence of the above assumptions is that one can incorporate information about the classification
of Lorentzian Lie algebras into the solution of the KSEs in a natural way. To illustrate this, supersymmetric
backgrounds with compact holonomy admit several $\hat\nabla$-parallel vector fields one of which is time-like constructed
from parallel spinor bi-linears.
If $dH=0$, one can show that their commutator is also $\hat\nabla$-parallel. Then an argument based on holonomy reduction
requires that the vector space spanned by the vector fields constructed from parallel spinor bi-linears
closes under Lie brackets. Since it is a Lorentzian Lie algebra, the classification results can be used
to determine the isometries of the supersymmetric backgrounds and as a result the geometry of spacetime.
We find that the number of Killing spinors $N$ depends on the
Lorentzian Lie algebra which acts on the spacetime.

Assuming that the action of the Lorentzian Lie algebra on the spacetime $M$ can be integrated to a free
group action, $M$  is a principal bundle
$M=P(G,B;\pi)$ with fibre group $G$ which generates the isometries of spacetime and base space $B$. Moreover,
it is equipped with a principal bundle connection, $\lambda$, which twists the fibre $G$ over $B$ \cite{het1}.
Using these data a brief description of the results we find after solving the KSEs is as follows.
In the $G_2$ case $(L=2)$,  there are solutions to the KSEs with 1 and 2 supersymmetries, see also table 4.
$\mathfrak{Lie}\,G$
as well as some of the properties of the dilaton for these solutions are summarized in table 1. The base space $B$
is a 7-dimensional manifold with a $G_2$ structure which is compatible with a connection with skew-symmetric torsion.
The connection $\lambda$ is a $G_2$ instanton with gauge group $G$.
In the $SU(3)$ case $(L=4)$, there are solutions with 1, 2 and 4 supersymmetries. $\mathfrak{Lie}\,G$ as well
as some of the geometric properties  of the base space are summarized in table 2. $B$ is a 6-dimensional manifold
equipped with either a $SU(3)$ or a $U(3)$
structure compatible with a connection with skew-symmetric torsion. It may also be an almost complex or complex manifold
depending on $G$ and $N$. The connection $\lambda$ is either a $SU(3)$ or $U(3)$ instanton, ie Hermitian-Einstein
connection\footnote{In the $U(3)$ case, there is a ``cosmological constant'' contribution along the diagonal $U(1)$ subgroup.},
with gauge group $G$. In the $SU(2)$ case $(L=8)$, there are solutions with 2, 4, 6 and 8 supersymmetries.
$\mathfrak{Lie}\,G$ as well
as some of the geometric properties  of the base space are summarized in table 3. $B$ is  a 4-dimensional manifold
which admits either a hyper-K\"ahler structure, or a K\"ahler structure or a the Weyl tensor is anti-self-dual. The connection $\lambda$
is either an anti-self-dual instanton, or a $U(2)$ instanton,  or the self-dual
part  satisfies a Hermitian-Einstein type of condition, with gauge group $G$.

This paper is organized as follows. In section 2, we review some aspects of the KSEs of heterotic supergravity and
specify the part of the dilatino KSE that we solve later. In section 3, we solve the KSEs for backgrounds
with holonomy $G_2$. In section 4, we solve the KSEs for backgrounds
with holonomy $SU(3)$. In section 5, we solve the KSEs for backgrounds
with holonomy $SU(2)$, and in section 6 we give our conclusions. In appendix A, we describe the solution
of the dilatino KSE for group manifolds.

\newsection{Gravitino and dilatino KSEs revisited}

Before we proceed to investigate each case separately, we shall first explain the general
characteristics of all cases and the strategy we have used to reformulate and  solve the dilatino KSE.
We  summarize some aspects of the gravitino KSE. The solution of the gaugino KSE remains unaltered and can be found
in \cite{het3}, see also \cite{eugauge}.

\subsection{Gravitino KSE}

The gravitino KSE of the heterotic supergravity has been solved in \cite{het1, het2}.
The spacetime of all supersymmetric backgrounds which admit $L$ parallel spinors with  compact isotropy
 group $K$ in $Spin(9,1)$, ${\rm hol}(\hat\nabla)\subseteq K$,
admits a local frame $e^A=(e^a, e^i)$ such that the spacetime metric $g$ and $H$ can be written as
\bea
ds^2&=&\eta_{ab}\, e^a e^b+\delta_{ij}\, e^i e^j~,
\cr
H&=& {1\over3!} H_{abc}\, e^a\wedge e^b\wedge e^c+{1\over2} H_{abi}\, e^a\wedge e^b\wedge e^i+{1\over2} H_{aij}\,
e^a\wedge e^i\wedge e^j+\tilde H~,
\la{gh}
\eea
respectively, where
\bea
\tilde H={1\over3!} H_{ijk}\, e^i\wedge e^j\wedge e^k~,
\eea
and  $\eta^{ab}=g^{-1}(e^a, e^b)$.
 The range of the indices $a$ and $i$ depends on the choice of
$K$.
Moreover the gravitino KSE implies that the forms
\bea
e^a~,~~~\tau\equiv {1\over k!}\tau_{i_1i_2\dots i_k} e^{i_1}\wedge e^{i_2}\wedge \dots\wedge e^{i_k}~,
\la{par}
\eea
are $\hat\nabla$-parallel, ie
\bea
\hat\nabla e^a=0~,~~~\hat\nabla\tau=0~,
\la{gravcon}
\eea
where $\tau$ stands for the fundamental
forms of $K$. Moreover one of the forms $e^a$ is time-like and all are constructed from
$\hat\nabla$-parallel spinor bi-linears.

The first condition in (\ref{gravcon}) implies that
\bea
{\cal L}_a g=0~,~~~~de^a=\eta^{ab}i_bH~,
\la{lah}
\eea
where the vector fields $e^M_a=g^{MN} \eta_{ab} e^b_N$ are dual to the 1-forms $e^a$ with respect to the spacetime metric $g$.
The second condition in (\ref{lah}) determines the $i_aH$ components of $H$ in terms of the exterior derivative of $e^a$.
Moreover the second condition in (\ref{gravcon}) can be solved to determine the rest of the components of
$H$ in (\ref{gh}) in terms of the spacetime metric
and $e^a$ and $\tau$ and their exterior derivatives.  In addition, it leads in some cases to set of conditions
on the $e^a$'s and  $\tau$'s which are interpreted as restrictions on the geometry of spacetime.
 This is the full content of the gravitino Killing spinor equation, see \cite{het1, het2} for more details.

{}For later use note that if $dH=0$, the Bianchi identity for $\hat\nabla$ yields
\bea
\hat R_{A[B,CD]}=-{1\over3} \hat\nabla_A H_{BCD}~.
\la{bbb}
\eea

\subsection{Dilatino KSE}

Before we proceed to organize the conditions that arise from the dilatino KSE, first observe that
$H$ in (\ref{gh}) and all tensors of $M$  can be decomposed further into irreducible representations of $K$.
 This is because $K$, as containing
the holonomy group of $\hat\nabla$, acts of the typical fibre of $TM$, the tangent space of spacetime,
with some representation. In particular, the typical fibre decomposes as $\bR^{9,1}=\bR^{9-\ell, 1}\oplus \bR^\ell$,
where $\bR^{9-\ell, 1}$ is spanned by the
directions of the parallel vector fields $e_a$ and the orthogonal complement
is taken with respect to the spacetime metric. $K$ acts trivially on $\bR^{9-\ell, 1}$ and with an irreducible
representation on $\bR^\ell$. Moreover since $K\subset Spin(\ell)\subset  Spin(9,1)$, its Lie algebra $\mathfrak{k}\subset \mathfrak{spin}(\ell)
=\Lambda^2(\bR^\ell)$. As a result, one can decompose $\Lambda^2(\bR^\ell)$ as $\Lambda^2(\bR^\ell)=\mathfrak{k}
\oplus \mathfrak{k}^\perp$. In turn, this will lead to a decomposition of the 2-forms on $M$. Similarly,
all tensors on $M$ can be decomposed in representations of $K$.

A direct inspection of the conditions which arise in the dilatino KSE for the descendants in \cite{het2}
reveals that they depend
on the tensors
\bea
[e_a, e_b]~,~~~(\tilde d e^a)\vert_{ \mathfrak{k}^\perp}~,~~~\partial_a\Phi~,~~~S~,
\la{apar}
\eea
where $\tilde d$ is the exterior derivative projected along the $e^i$ directions, $S$ is the singlet
in the decomposition of $\tilde H$ under $K$ and $(\tilde d e^a)\vert_{ \mathfrak{k}^\perp}$ denotes
the projection of the 2-form $\tilde de^a$ on $\mathfrak{k}^\perp$.

On the other hand under the  assumptions, $dH=0$ and ${\rm hol}(\hat\nabla)\subseteq K$, an analysis of the
Bianchi identity  (\ref{bbb}) for the
$\hat\nabla$ connection reveals that
 the tensors (\ref{apar}) are $\hat\nabla$-parallel.
Now if the forms (\ref{apar}) are linearly independent from those of (\ref{par}), then
${\rm hol}(\hat\nabla)$ reduces to a proper subgroup of $K$. We shall explain in each case later
that the pattern of reduction is
\bea
G_2\Longrightarrow SU(3)\Longrightarrow SU(2)\Longrightarrow \{1\}~.
\eea
So if we assume that (\ref{apar}) are linearly independent from those (\ref{par}) for a group $K$,
then simply we have to consider
the next case on  with more parallel spinors. Since the sequence terminates,  to examine all cases it suffices
to take (\ref{apar}) to be linearly dependent on (\ref{par}). As a result, one has
\bea
\partial_a\Phi={\rm const}~,~~~[e_a, e_b]=- H_{ab}{}^c e_c~,~~~(\tilde de^a)\vert_{\mathfrak{k}^\perp}
= f^a\, \tau~,~~~S=\nu\,\tau~,
\la{lin}
\eea
where $f^a$ and $\nu$ are constants  and the last two equations are schematic. Therefore it is required that
the vector space spanned by vector fields constructed as
spinor bi-linears closes under Lie brackets and so it is a Lorentzian Lie algebra. Thus, in particular,
$[e_a, e_b]_i=0$ which implies that
\bea
H_{abi}=0~.
\la{habi}
\eea
This leads to an extensive simplification of the conditions that arise from the dilatino KSE. Moreover,
 the Lorentzian Lie algebras that arise in each case have been
classified and they will be given for each case separately.

Using (\ref{lin}) and (\ref{habi}), a direct inspection of the conditions that arise from the dilatino
Killing spinor equation in \cite{het2},
reveals  that it factorizes\footnote{This factorization is closely related to the decomposition
of the dilatino KSE into the irreducible representation of $K$ on $\bR^\ell$ and the singlets.}. One part gives the  condition
\bea
\theta_\tau=2\tilde d \Phi~,
\la{leee}
\eea
where $\theta_\tau$ is the Lee form of one of the fundamental forms of $K$. The expressions of all
the relevant Lee forms can be found at \cite{het1, het2}.
This  encompasses  the contribution that
 $\tilde d\Phi$ and  $\tilde H$, apart from the singlet $S$, make
in the dilatino KSE.

The other part of  the dilatino KSE involves always  the tensors (\ref{apar}),
and
 can be written as
\bea
(\Gamma^a\partial_a\Phi-{1\over12} H_{abc}\Gamma^{abc}-{1\over12}S_{ijk}\Gamma^{ijk}-{1\over 4} H_{aij}
 \Gamma^{aij})\epsilon=0~.
\la{sd}
\eea
Observe that in the last term  only the $\mathfrak{k}^\perp$ component of $\tilde de^a$ contributes
because the spinors $\epsilon$ are $K$-invariant and so are annihilated by the $\mathfrak{k}$ part.

The condition (\ref{leee}) is known in all cases. So it remains to solve (\ref{sd}).  The $K=G_2$ case is simple
enough to incorporate the data (\ref{apar}) in the calculation of \cite{het2} without having to examine (\ref{sd}).
For $K=SU(3)$ and $SU(2)$,
we shall use the classification of Lorentzian Lie algebras \cite{medina, josek} to specify the structure constants $H_{abc}$
and then we shall proceed to solve (\ref{sd}). This is achieved by writing (\ref{sd}) as a sum
of commuting operators acting on the $K$-invariant spinors $\e$ and by analyzing their eigen-values and eigen-spaces.
The analysis for $K=\{1\}$ has been done in \cite{josek}, see also \cite{het2}.

{}We have argued that the vector space spanned by the parallel vector bi-linears is closed under Lie brackets
and so can be identified with a Lie algebra $\mathfrak{Lie}(G)$ of a group $G$. Moreover ${\cal L}_aH=0$
since $i_aH=de^a$ and $dH=0$.
Assuming that the infinitesimal group action generated by the vector field $e_a$ can be integrated into a free
group action
on the spacetime\footnote{$e_a$'s  never vanish since they are parallel. So the assumption involves
the appropriate closure of the orbits generated by the vector fields.}, the spacetime is a principal bundle $M=P(G,B;\pi)$ with base space $B$ such that
\bea
ds^2&=&\mt_{ab} \lambda^a \lambda^b+\pi^*d\tilde s^2
\cr
H&=&{1\over6} H_{abc} \lambda^a\wedge \lambda^b\wedge \lambda^c+\mt_{ab} \lambda^a\wedge {\cal F}^b+\pi^* \tilde H~,
\la{metreH}
\eea
where $\lambda^a\equiv e^a$ is a principal bundle connection and
\bea
{\cal F}^a\equiv \tilde d e^a:=d\lambda^a-{1\over 2} H^a{}_{bc}\lambda^b\wedge \lambda^c\equiv{1\over2} H^a{}_{ij} e^i\wedge e^j~,
\eea
is the associated curvature.
 $H$ is the sum of the Chern-Simons form of
$\lambda$ and a 3-form $\tilde H$ of $B$.

Therefore to specify the geometry of the spacetime in each case, one has to describe three pieces of data, (i)
the Lorentzian Lie algebra $\mathfrak{Lie}(G)$, (ii) the geometry of the base space $B$ with respect
to the pair $(d\tilde s^2, \tilde H)$, and (iii)
how the fibre $G$ twists over the base space. The latter is determined by the conditions that the curvature
${\cal F}$ of the principal bundle connection satisfies. For the field equations of the theory
and details about the notation see \cite{het1, het2}.

\newsection{$G_2$}

\subsection{ Holonomy reduction}

The backgrounds with ${\rm hol}(\hat\nabla)\subseteq G_2$  admit three 1-forms $e^a$ that can be constructed from
 Killing spinor bilinears. Therefore the typical
fibre of $TM$ decomposes as $\bR^{2,1}\oplus \bR^7$. Moreover $G_2$ acts on $\bR^7$ with the 7-dimensional
irreducible representation. The isotropy group in $G_2$ of a vector in $\bR^7$  is $SU(3)$.  Using the Bianchi
identity (\ref{bbb}) and $dH=0$, one can show that $[e_a, e_b]$ are $\hat\nabla$-parallel \cite{het2}. Thus
if $[e_a, e_b]$ are linearly independent from $\{e_a\}$, then the holonomy of $\hat\nabla$ reduces
to a subgroup of $SU(3)$.  This is a special case of backgrounds with $SU(3)$ holonomy that will be
investigated later. Thus to examine a $G_2$ case that does not reduce to an $SU(3)$ one, we shall take
that the vector space spanned by the $e_a$'s closes under Lie brackets and becomes 3-dimensional Lorentzian
Lie algebra. These have been classified and are isomorphic to
\bea
\bR^{2,1}~,~~~\mathfrak{sl}(2, \bR)~.
\eea

Next $\Lambda^2(\bR^7)=\mathfrak{g}_2\oplus \bR^7$. Thus $(de^a)\vert_{\mathfrak{g}_2^\perp}$
are also in the 7-dimensional
representation of $G_2$. Again the Bianchi identity  (\ref{bbb}) and $dH=0$ imply that $(de^a)\vert_{\mathfrak{g}_2^\perp}$
is $\hat\nabla$-parallel. Thus if it does not vanish, then the holonomy
group again reduces to $SU(3)$.  Thus to investigate the $G_2$ backgrounds which do not reduce
to those of $SU(3)$, we must take
\bea
(de^a)\vert_{\mathfrak{g}_2^\perp}=0~.
\eea

Now there is only one singlet representation of $G_2$ in $\Lambda^3 (\bR^7)$ and this is proportional to the
fundamental $G_2$ 3-form $\varphi$. Thus we write
\bea
S_{ijk}=\nu\, \varphi_{ijk}~.
\eea
Moreover $\nu$ is not arbitrary but rather
\bea
\nu=
-{1\over6} (\tilde d\varphi, \star\varphi)~,
\eea
since all $\tilde H$ is expressed in terms of the fundamental forms as a consequence of the
gravitino KSE\footnote{ In particular, $
\tilde H=-{1\over6} (\tilde d\varphi, \star\varphi) \varphi+ \star \tilde d\varphi-\star(\theta_\varphi\wedge \varphi)
$, see \cite{ivanov}.}.
Thus for the $G_2$ case, equation (\ref{lin}) is written as
\bea
\partial_a\Phi={\rm const}~,~~~[e_a, e_b]=- H_{ab}{}^c e_c~,~~~(\tilde de^a)\vert_{\mathfrak{g}_2^\perp}=0~,~~~
S_{ijk}=\nu\, \varphi_{ijk}~,
\la{g2lin}
\eea
where the structure constants $H_{abc}$ either vanish or are those of $\mathfrak{sl}(2,\bR)$.

\subsection{Dilatino KSE}

The (\ref{leee}) part of the dilatino KSE is
\bea
\theta_\varphi=2\tilde d \Phi~.
\eea
Using (\ref{g2lin}), (\ref{sd})  becomes
\bea
(\Gamma^a\partial_a\Phi-{1\over12} H_{abc}\Gamma^{abc}-{1\over12} \nu \varphi_{ijk} \Gamma^{ijk})\epsilon=0~,
\la{g2sd}
\eea
where $H_{abc}$ are the structure constants of either $\bR^{2,1}$ or $\mathfrak{sl}(2, \bR)$. To proceed,
we shall examine this equation for $\bR^{2,1}$ and $\mathfrak{sl}(2,\bR)$ separately.

\subsubsection{$\bR^{2,1}$}

In the $\bR^{2,1}$ case  $H_{abc}=0$. First suppose that $\nu=0$. In such case,
 (\ref{g2sd}) has a solution iff
\bea
\eta^{ab}\partial_a\Phi \partial_b\Phi=0~.
\eea
Thus there are backgrounds preserving one supersymmetry, $N=1$, provided that $\partial_a\Phi$ spans a
non-vanishing null direction in the Lie algebra $\bR^{2,1}$. Therefore the dilaton
is linear along a light-cone fibre coordinate $z^+$ but in general
depends non-linearly on the coordinates $x$ of the base space $B$, ie
\bea
\Phi=c z^++ b(x)~.
\eea
However if $\partial_a\Phi=0$, then the dilatino
KSE vanishes identically and such backgrounds preserve two supersymmetries. The dilaton is
constant along the fibre directions but again depends non-linearly on the coordinates of the base space.

Next suppose that $\nu\not=0$. A direct inspection of the results of \cite{het2} reveals that there is an $N=1$
solution in this case provided that the dilaton is linear along a space-like fibre direction.
Take $\partial_1\Phi\not=0$,
then
\bea
\partial_1\Phi={7\over2}\nu=-{7\over12} (\tilde d\varphi, \star\varphi)
\eea

If the dilaton is constant along the fibre directions, ie $\partial_a\Phi=0$,  then $\nu=0$ and the
 backgrounds preserve 2 supersymmetries.

\subsubsection{$\mathfrak{sl}(2, \bR)$}
If  $\mathfrak{Lie}\,G=\mathfrak{sl}(2, \bR)$,  the field equation of the 3-form flux implies that
\bea
\partial_a\Phi H^a{}_{bc}=0~.
\eea
Thus the $\partial_a\Phi$ direction in  $\mathfrak{sl}(2, \bR)$ commutes with all the others. Since
 $\mathfrak{sl}(2, \bR)$ is simple, there is no such direction and so it is required that
\bea
\partial_a\Phi=0~.
\eea
Thus all backgrounds with $SL(2, \bR)$ fibre have constant dilaton along the fibre direction. Of course
the dilaton still depends on the coordinates on $B$.
 A direct inspection of the results of \cite{het2} reveals that all such backgrounds  preserve 2 supersymmetries\footnote{
 Solutions of the heterotic KSEs containing $SL(2,\bR)$ have been emphasized in \cite{japan}.}.
 Moreover
\bea
7 (\tilde d\varphi, \star\varphi)=H_{-+1}~,
\eea
where $H_{-+1}$ are the structure constants of $\mathfrak{sl}(2,\bR)$.
The results are summarized in table 1.

\begin{table}[ht]
 \begin{center}
\begin{tabular}{|c|c|c|c|}\hline
$G/d\Phi$&${\rm spacelike}$&${\rm null}$&${\rm zero}$
 \\
\hline\hline
 $\bR^{2,1}$&$1$  & $1$ & $2$\\
 \hline
$\mathfrak{sl}(2,\bR)$&$-$&$-$ &$2$\\
\hline
\end{tabular}
\end{center}
\caption{\small
The entries $-$ do not occur. The terms spacelike, null
and zero are referred to $d\Phi$ along the group fibre directions. The numerical entries are the number
of supersymmetries preserved in each case.}
\end{table}

\subsection{Geometry}

As it has already been mentioned, the spacetime is a principal bundle with fibre $\bR^{2,1}$ or $SL(2,\bR)$. Having
specified the geometry of the fibre, it remains to find the geometry of the base space $B$ and how the fibres
twist over the base space.

First let us begin with the geometry of the base space. Since  for all backgrounds
$\tilde de^a\vert_{\mathfrak{g}_2^\perp}=0$, as  required for the holonomy group not to reduce to $SU(3)$, the fundamental
forms of $G_2$ satisfy
\bea
i_a\varphi=i_a\star\varphi=0~,~~~{\cal L}_a \varphi={\cal L}_a \star\varphi=0~,
\eea
where $\star$ is the Hodge operation of the directions spanned by the $e^i$'s.
Therefore both $\varphi$ and $\star\varphi$ descent on the base space $B$, and so $B$ has a $G_2$-structure. This structure
is compatible with a
connection with skew-symmetric torsion $\hat\tilde\nabla$ given by the data $(d\tilde s^2, \tilde H)$,
where $d\tilde s^2=
\delta_{ij} e^i e^j$. Such geometries have been investigated in \cite{ivanov, het1, het2}.

The dilatino KSE requires the additional condition
\bea
(\tilde d\varphi, \star\varphi)=0~,
\eea
for the cases $\bR^{2,1}$ $N=2$, and $\bR^{2,1}$ $N=1$ when the dilaton is null. In the rest of the cases
$(\tilde d\varphi, \star\varphi)\not=0$ and it is related either to the structure constants of $\mathfrak{sl}(2,\bR)$
or to the spacelike linear dilaton along the group fibres for the $\bR^{2,1}$, $N=1$ case.

The fibre twists over the base space with a principal $\bR^{2,1}$ or $SL(2,\bR)$ connection which is a $G_2$-instanton.
This is because $\tilde de^a\vert_{\mathfrak{g}_2^\perp}=0$ and so
\bea
{\cal F}^a\vert_{\mathfrak{g}_2^\perp}=0~.
\eea

Therefore the spacetime can be reconstructed in all cases starting from a 7-dimensional $G_2$-manifold $B$
compatible with a metric connection with skew-symmetric torsion and a  $G_2$-instanton connection over $B$
with gauge group $\bR^{2,1}$ or $SL(2,\bR)$. It is characteristic that all backgrounds with $N=1$ supersymmetry
have a linear dilaton along the fibre group directions.

\newsection{$SU(3)$}

\subsection{Holonomy Reduction}

The backgrounds with ${\rm hol}(\hat\nabla)\subseteq SU(3)$ admit four 1-forms $e^a$ constructed from
parallel spinor bi-linears \cite{het1}.  So the typical fibre of $TM$ decomposes as $\bR^{9,1}=\bR^{3,1}\oplus \bR^6$ and
$SU(3)$ acts on $\bR^6\otimes \bC=\bC^3\oplus \bar\bC^3$ with the fundamental representation and its complex conjugate.
Moreover a vector in $\bC^3$ (or $\bar\bC^3$) has isotropy group $SU(2)$ in $SU(3)$. As in the $G_2$ case, the Bianchi
identity (\ref{bbb}) and $dH=0$ imply that $[e_a, e_b]$ is $\hat\nabla$-parallel. Thus if one
of the vector fields $[e_a, e_b]$ is linearly independent from those in $\{e_a\}$, then the holonomy
of $\hat\nabla$ reduces to a subgroup of $SU(2)$. Such backgrounds  are included in those with holonomy $SU(2)$
which will be examined  in the next section. To investigate $SU(3)$ backgrounds which
do not reduce to $SU(2)$ ones,  we shall take that the vector space spanned by $\{e_a\}$  closes
under Lie brackets. So  $\{e_a\}$'s span a 4-dimensional Lorentzian Lie algebra and these are isomorphic to
\bea
\bR^{3,1}~,~~~\mathfrak{sl}(2,\bR)\oplus \bR~,~~~\bR\oplus \mathfrak{su}(2)~,~~~\mathfrak{cw}_4~.
\la{su3a}
\eea

The fundamental forms of $SU(3)$ are the Hermitian 2-form $\omega$ and the holomorphic volume 3-form $\chi$ which
are chosen
as
\bea
\omega=-e^2\wedge e^7- e^3\wedge e^8-e^4\wedge e^9~,~~~\chi=(e^2+ie^7)\wedge (e^3+ie^8)\wedge (e^4+i e^9)~,
\eea
and so $a,b=0,5,1,6$.
To
identify the components of $\tilde d e^a$ along $\mathfrak{su}(3)^\perp$,  we decompose $\tilde d e^a$
 in (2,0), (1,1) and (0,2)-forms  using  the almost complex structure constructed
from the fundamental form $\omega$ of $SU(3)$ and $d\tilde s^2$. The directions that lie along $\mathfrak{su}(3)^\perp$ are the
(2,0) and (0,2) components as well as the (1,1) component along the Hermitian form $\omega$.

The (0,2) and (2,0) components of $(\tilde e^a)\vert_{\mathfrak{su}(3)^\perp}$  lie in the fundamental
representation of $SU(3)$
and its complex conjugate. Again the Bianchi identity  (\ref{bbb}) and $dH=0$ imply that they are $\hat\nabla$-parallel.
Thus if they do not vanish, then the holonomy of these $SU(3)$  backgrounds reduces
to $SU(2)$. So to investigate  $SU(3)$ backgrounds which are not special cases
of those with holonomy $SU(2)$, we have to set
\bea
(\tilde d e^a)^{2,0}=0~;
\eea
the (0,2) component is complex conjugate to (2,0). The component of $(\tilde d e^a)\vert_{\mathfrak{su}(3)^\perp}$
along $\omega$ does not reduce the holonomy of the spacetime because it is proportional to
a fundamental form of $SU(3)$. Thus we can set
\bea
(\tilde d e^a)\vert_{\mathfrak{su}(3)^\perp}=f^a \omega~,
\la{su3fo}
\eea
where $f$ is a constant.

It remains to identify the singlet component $S$ of $\tilde H$. In the decomposition of
$\Lambda^3(\bR^6)\otimes\bC$ under $SU(3)$ there is a unique singlet representation proportional to
the fundamental form $\chi$ and its complex conjugate. Thus we can set
\bea
S= {1\over 2\sqrt{2}}\,(\mu\, \chi+\bar\mu\, \bar\chi)~.
\eea
The Bianchi identity (\ref{bbb}) and $dH=0$ imply that $\mu$ is a complex constant,
 and the normalization numerical factor has been chosen for convenience.

 Therefore,  the equation (\ref{lin}) for the $SU(3)$ case can be written as
\bea
\partial_a\Phi={\rm const}~,~~[e_a, e_b]=- H_{ab}{}^c e_c~,~~
(\tilde d e^a)\vert_{\mathfrak{su}(3)^\perp}= f^a \omega~,~~S= {1\over 2\sqrt{2}}\,(\mu\, \chi+\bar\mu\, \bar\chi)~,
\la{su3lin}
\eea
where the structure constants  $H_{abc}$ are those of one of the Lie algebras in (\ref{su3a}).

Before we proceed to examine the dilatino KSE case by case, we shall establish that
\bea
\partial_a\Phi\, H^a{}_{bc}=0~,~~~f_a\, H^a{}_{bc}=0~.
\la{phfh}
\eea
The first follows from the field equation of 2-form gauge potential as in the $G_2$ case. To establish the latter
equation, let us compute the Lie derivative of $\chi$ along the Killing vector direction $e_a$ and use (\ref{su3fo})
to find
\bea
{\cal L}_a\chi= -3i f_a \chi~.
\la{su3lc}
\eea
Since $[{\cal L}_X, {\cal L}_Y]={\cal L}_{[X,Y]}$, it is easy to see that consistency requires the second equation
in (\ref{phfh}). Therefore the directions spanned by $\partial_a\Phi$ and $f$ are central in $\mathfrak{Lie}\,G$.

\subsection{Dilatino KSE}

The (\ref{leee}) part of the dilatino KSE is
\bea
\theta_\omega=2\tilde d \Phi~,
\eea
and the (\ref{sd}) can be written as
\bea
\big(\Gamma^a\partial_a\Phi-{1\over12} H_{abc}\Gamma^{abc}-{1\over 4}f_a \omega_{ij}  \Gamma^{aij}-
{1\over2} (\mu\, \Gamma^{234}+\bar\mu\, \Gamma^{\bar2\bar3\bar4})\big)\epsilon=0~,
\la{su3sd}
\eea
where we have expressed the singlet part $S$ of $\tilde H$ explicitly in terms of  Hermitian  gamma matrices.

The simplest case to consider first is $f=\mu=0$. The dilatino KSE depends only on the dilaton and the
structure constants $H_{abc}$. This is the case examined in \cite{josek}. Here the dilatino KSE
 has been analyzed in detail
for the holonomy $SU(2)$ case in appendix A because it is more delicate. The analogous $SU(3)$ case is
straightforward. The solutions
preserve either  2 or all 4 supersymmetries.

\subsubsection{$\bR^{3,1}$}

\vskip 0.2cm
\underline {$N=1$}
\vskip 0.2cm
In this case $H_{abc}=0$. Next suppose that $\partial_a\Phi\not=0$. Multiplying  (\ref{su3sd}) with
$\partial_b\Phi\Gamma^b$, we find that it can be written as
\bea
[(\partial_a\Phi)^2-{1\over2} \partial_a\Phi f^a c(\omega)-{1\over4} \partial_a\Phi f_b c(\omega)
\Gamma^{ab}-{1\over2} \partial_a\Phi \Gamma^a c(S)]\e=0~,
\la{su3sd1}
\eea
where
\bea
c(\omega)={1\over2} \omega_{ij} \Gamma^{ij}~,~~~c(S)=\mu\, \Gamma^{234}+\bar\mu\, \Gamma^{\bar2\bar3\bar4}~.
\eea
Observe that on $SU(3)$-invariant spinors $[c(\omega)]^2=-9\, 1_{4\times4}$ and $[c(S)]^2=-8\mu \bar\mu \, 1_{4\times4}$.
Then define the traceless matrices
\bea
A={1\over2} \partial_a\Phi f^a c(\omega)+{1\over2} \partial_a\Phi \Gamma^a c(S)~,~~~B={1\over4} \partial_a\Phi f_b c(\omega)
\Gamma^{ab}
\eea
and observe that on $SU(3)$-invariant spinors
\bea
A^2= \Delta_1^2\, {\rm Id}~,~~~B^2= \Delta_2^2\, {\rm Id}~,~~~AB=BA~,
\eea
where
\bea
\Delta_1^2&=&-{9\over4} (\partial_a\Phi f^a)^2+2\mu\bar\mu \partial_a\Phi \partial^a\Phi~,
\cr
 \Delta_2^2&=& {9\over4} [\partial_a\Phi \partial^a\Phi f_b f^b- (\partial_a\Phi f^a)^2]~.
\eea

Now there are various possibilities to consider. First observe that if $\partial_a\Phi \partial^a\Phi\not=0$,
multiplying the (\ref{su3sd}) with $\partial_a\Phi\Gamma^a$ is an invertible operation and so (\ref{su3sd1})
is equivalent to (\ref{su3sd}). The null  $\partial_a\Phi \partial^a\Phi=0$ and $\partial_a\Phi=0$ cases
will  be investigated separately, and both of them give backgrounds which preserve at least 2 supersymmetries.

So let us take $\partial_a\Phi \partial^a\Phi\not=0$. Next if
\bea
\Delta_1^2, \Delta_2^2>0
\eea
one can decompose the $SU(3)$-invariant spinors in eigen-spaces of $A$ and $B$.  This in particular implies
that $\partial_a\Phi$ and $f_a$ are spacelike and non-co-linear.  $N=1$
supersymmetric backgrounds exist provided that
\bea
(\partial_a\Phi)^2\mp \Delta_1\mp \Delta_2=0
\la{su3dd}
\eea
the relative ambiguous signs are uncorrelated. For example, $N=1$ solutions exist if both $\partial_a\Phi$ and $f_a$
are orthogonal and space-like.

Next if either
\bea
\Delta_1^2<0~,~~{\rm or}~~~\Delta_2^2<0~,
\eea
 then there are no solutions. This is because
the spinors in the heterotic case are real and either $A$ or $B$ can be diagonalized only over complex
spinors.

There are two marginal cases to consider depending on whether either $\Delta_1^2$ or $\Delta_2^2$ vanish. In such case
either $A$ or $B$ become nilpotent. Notice that $f$ and $\partial_a\Phi$ are either space-like or null. Since the null
case will be investigated later, we take  $(\partial_a\Phi)^2>0$.
To begin, suppose that $A$ is nilpotent, $A^2=0$, ie
\bea
\Delta_1=0~,~~~\Delta_2^2>0~.
\eea
Decomposing the Killing spinor in eigenvalues of $B$ as $B\epsilon_\pm=\pm \Delta_2 \epsilon_\pm$, $\Delta_2>0$,
and since
$A$ commutes with $B$, the dilatino KSE becomes
\bea
((\partial_a\Phi)^2-\Delta_2) \epsilon_+-A\epsilon_+&=&0
\cr
((\partial_a\Phi)^2+\Delta_2) \epsilon_--A\epsilon_-&=&0
\eea
Acting  on the second equation with $A$ and using $A^2=0$, we find that $A\epsilon_-=0$. Substituting back,
 we arrive at
$\epsilon_-=0$.  Similarly acting on the first equation with $A$ and using a similar argument, we conclude
that $\epsilon_+=0$ unless $(\partial_a\Phi)^2-\Delta_2=0$, ie that (\ref{su3dd}) is valid in the limit $\Delta_1=0$
for one choice of sign. Now substituting this into the first equation above, we find that $A\epsilon_+=0$ as well.
It remains to find the dimension of the kernel of $A$ on the eigenspace of $B$ with  eigenvalue $\Delta_2$. For this
first observe that $\partial_a\Phi f^a\not=0$ since otherwise $\Delta_1=0$ would imply that $\partial_a\Phi$
is null. Using this, the condition
 $A\epsilon_+=0$ can be rewritten as
\bea
\big [1-{\partial_a\Phi\Gamma^a c(\omega) c(S)\over 9\, \partial_a\Phi f^a}\big ]\epsilon_+=0
\eea
Since the second term is traceless and has eigenvalues $\pm 1$ for $\Delta_1=0$, the kernel is 1-dimensional. These
solutions are included in   (\ref{su3dd}) at the limit that $\Delta_1$ vanishes.

Next take
\bea
\Delta_1^2>0~,~~~\Delta_2=0~,
\eea
in which case $B^2=0$.
As in the previous case, separating the dilatino KSE in eigenspaces, $\epsilon_\pm$, of $A$ with eigenvalues
 $\pm \Delta_1$,
we find that the dilatino KSE is satisfied provided that $\epsilon_-=0$, $B\e_+=0$,  and
(\ref{su3dd}) is satisfied for $\Delta_2=0$ for one choice of sign. It remains to determine the kernel of $B$.
For this, write $f_a=\ell \partial_a\Phi+u_a$, where $\partial_a\Phi u^a=0$, $u\not=0$, and $\ell$ is a constant.
As in the previous case, we take $(\partial_a\Phi)^2>0$ and so $\Delta_2=0$ implies that $u$ is null.
In terms of $u$
$B$ is written as
\bea
B={1\over4} \partial_a\Phi u_b c(\omega) \Gamma^{ab}~.
\eea
Therefore $B\e_+=0$ implies that
\bea
u_a\Gamma^a\e_+=0
\eea
and so $\e$ satisfies the standard chiral projection condition.

Now suppose that $\partial_a\Phi$ is null. In such case $\Delta^2_2< 0$ unless $\partial_a\Phi f^a=0$. Thus
in the null case $\Delta_1=\Delta_2=0$. Assume that $f\not=0$ and act on (\ref{su3sd}) with $f_a\Gamma^a c(\omega)$.
This gives
\bea
[f_a\partial_b\Phi \Gamma^{ab} c(\omega)+{9\over2} f^2-{1\over2} f_a\Gamma^a c(\omega) c(S)]\e=0~.
\la{su3sd2}
\eea
Set $L={1\over2} f_a\Gamma^a c(\omega) c(S)$ and $K=f_a\partial_b\Phi \Gamma^{ab} c(\omega)$. Now
\bea
L^2=18 f^2 {\rm Id}~,~~~K^2=0~,~~~LK=KL~,
\eea
Thus solutions exist iff $f^2\geq 0$. Moreover acting on (\ref{su3sd2}) with $K$, we find that
\bea
({9\over2} f^2-L) K\e=0~.
\eea
Since the eigenvalues of $L$ are different from ${9\over2} f^2$, the only solutions of this
are that either $K\e=0$  or $f^2=0$. In the former case substituting this back into (\ref{su3sd2}),
one concludes that $\e=0$ and so such backgrounds are not supersymmeric. In the latter case, both
$f_a$ and $\partial_a\Phi$ are null and co-linear. Thus acting on (\ref{su3sd}) with $\partial_a\Gamma^a$, one
finds that
\bea
\partial_a\Phi\Gamma^a c(S)\e=0
\eea
which in turn implies that $\partial_a\Phi\Gamma^a\e=0$, ie without loss of generality one can set
\bea
\Gamma^+\e=0~.
\eea
Substituting this back into (\ref{su3sd}), one concludes that there are solutions iff
\bea
\mu=0
\eea
ie the (3,0) part of $\tilde H$ vanishes. As we shall see this implies that $B$ is a complex manifold.
In addition observe  that if $f$ and $\partial_a\Phi$ are null and co-linear and $\mu=0$, then  the matrices $A=B=0$.
As  we shall explain such backgrounds
preserve at least 2 supersymmetries.

It remains to investigate the case that $\partial_a\Phi=0$. For this, it is easy to see that if $f^2=0$ as
well, the dilatino KSE has no solutions unless $\mu=0$. Thus to proceed, we take $f^2\not=0$. Acting with
$f_a \Gamma^a c(\omega)$ on the dilatino KSE and after some rearrangement, it can be rewritten
as
\bea
\big [1-{f_a\Gamma^a c(\omega) c(S)\over 9 f^2} \big ]\e=0~.
\eea
This has solutions provided that
\bea
f^2={8\over9}\mu\bar\mu~.
\eea
Moreover such backgrounds preserve at least 2 supersymmetries.
Of course if  $f=\partial_a\Phi=0$ and  $\mu=0$, then the solutions preserve 4 supersymmetries.

\vskip 0.2cm
\underline {$N=2$}
\vskip 0.2cm

We have already seen that if $\partial_a\Phi$ is null or zero, then the solutions preserve 2 supersymmetries.
Thus it remains to find the solutions of the Killing spinor equations provided that $(\partial_a\Phi)^2\not=0$
In such case, the solutions of the Killing spinor equations will preserve two supersymmetries iff either
$A$ or $B$ vanish identically. The possibility of both vanishing is included in the case that $\partial_a\Phi$
is null.

First for $A=0$, one has that $\partial_a\Phi f^a=0$ and $\mu=0$. There are solutions provided that
both $\partial_a\Phi$ and $f$ are spacelike and orthogonal, and (\ref{su3dd}) is satisfied.
Next $B=0$, iff $f_a=\ell \partial_a\Phi$, ie $f$ and $\partial_a\Phi$ are co-linear. Solutions
preserving 2 supersymmetries exist provided that $\Delta_2^2> 0$ and (\ref{su3dd}) is satisfied.
For these solutions $\mu\not=0$. There are no solutions preserving 3 supersymmetries.

\subsubsection{$\bR\oplus \mathfrak{su}(2)$}

Without loss of generality, we can assume that $\mathfrak{su}(2)$ spans the directions $1,5,6$. Eqn
(\ref{phfh}) implies that the only non-vanishing component of $\partial_a\Phi$ and $f_a$ is
$\partial_0\Phi$ and $f_0$. Setting
\bea
{1\over6} H_{abc}\Gamma^{abc}=\nu \Gamma^{516}
\eea
where $\nu$ is a constant, and acting with $\Gamma^{516}$  onto the KSE (\ref{sd}), one finds
\bea
[-\partial_0\Phi \Gamma^{0516}+{\nu\over2}+{f_0\over2} \Gamma^{0516}c(\omega)-{1\over2} \Gamma^{516} c(S)]\e=0~.
\eea
Next define
\bea
A={1\over2} f_0 \Gamma^{0516} c(\omega)~,~~~B=-\partial_0\Phi \Gamma^{0516}-{1\over2} \Gamma^{516} c(S)~,
\eea
and observe that
\bea
A^2={9\over4} f_0^2~,~~~B^2=-(\partial_0\Phi)^2-2\mu\bar\mu~,~~~AB=BA~.
\eea
If $(\partial_0\Phi)^2+2\mu\bar\mu\not=0$, the eigenvalues of $B$ are complex and there are no solutions.
So, we should take
\bea
\partial_0\Phi=0~,~~~\mu=0~.
\eea
Thus the dilaton is constant along the fibre directions and the Nijenhuis tensor of the base space $B$ vanishes.

In such case, the dilatino KSE becomes
\bea
[{\nu\over2}+{1\over2} \Gamma^{0516} c(\omega)]\e=0
\eea
So there are solutions   provided that
\bea
\nu=\pm {3} f_0~.
\eea
In fact, all solutions preserve 4 supersymmetries.

\subsubsection{$\mathfrak{sl}(2)\oplus \bR$}

Without loss of generality, let us assume that $\mathfrak{sl}(2)$ spans the directions $0, 5, 1$. In such case, eqn
(\ref{phfh}) implies that the only non-vanishing component of $\partial_a\Phi$ and $f_a$ is
$\partial_6\Phi$ and $f_6$. Next write
\bea
{1\over6} H_{abc} \Gamma^{abc}=\nu \Gamma^{051}
\eea
and multiply (\ref{su3sd}) with $\Gamma^{051}$ to find
\bea
[\partial_6\Phi \Gamma^{0516}-{1\over2} \nu -{1\over2} f_6 \Gamma^{0516} c(\omega)-{1\over2} \Gamma^{051} c(S)]\e=0~.
\eea
Then as in the previous case define
\bea
A=-{1\over2} f_6 \Gamma^{0516} c(\omega)~,~~~B=\partial_6\Phi \Gamma^{0516} -{1\over2} \Gamma^{051} c(S)~,
\eea
and observe that
\bea
A^2={9\over4} f_6^2~,~~~B^2=-(\partial_6\Phi)^2+2\mu\bar\mu~,~~~AB=BA~.
\eea
Now there are various cases to be considered. If $-(\partial_6\Phi)^2+2\mu\bar\mu<0$, the eigenvalues of $B$ are complex
and there are no solutions. On the other hand if $-(\partial_6\Phi)^2+2\mu\bar\mu>0$, there are solutions
preserving 1 supersymmetry provided that
\bea
-{1\over2} \nu\pm {3\over2} f_6\pm \sqrt{-(\partial_6\Phi)^2+2\mu\bar\mu}=0~,
\la{n1su3}
\eea
where the signs are uncorrelated. Notice that if $f_6=0$, the solutions preserve 2 supersymmetries.

It remains to investigate the case $(\partial_6\Phi)^2=2\mu\bar\mu$. Separating the dilatino KSE on the eigenspaces
of $A$ we have
\bea
[B-{1\over2} \nu \pm {3\over2} f_6]\e_\pm=0~.
\eea
Acting with $B$ and using $B^2=0$, there are solutions provided that
\bea
\nu=\pm 3 f_6~.
\eea
Choosing one of the signs, say the positive sign, the Killing spinor equation becomes
\bea
[\partial_6\Phi-{1\over2} \Gamma^6 c(S)]\e_+=0~.
\eea
Now $(\Gamma^6 c(S))^2=8\mu\bar\mu$, so there are solutions that preserve 1 supersymmetry provided
\bea
\partial_6\Phi\pm \sqrt {2 \mu\bar\mu}=0~.
\eea
Clearly, the $B^2=0$ case is included in that of (\ref{n1su3}) for special values of the parameters.

There are backgrounds with 2 supersymmetries if at least one of the operators $A$ or $B$ vanish identically.
$A$ vanishes if $f_6=0$ and as we have mentioned the solutions preserve 2 supersymmetries. On the other hand $B$
vanishes if both $\partial_6\Phi$ and $\mu=0$. In this case the solutions preserve 4 supersymmetries.

\subsubsection{$\mathfrak{cw}_4$}

Without loss of generality, let us assume that $\mathfrak{cw}_4$ spans the directions $+,-, 1,6$. In such case, eqn
(\ref{phfh}) implies that the only non-vanishing component of $\partial_a\Phi$ and $f_a$ is
$\partial_+\Phi$ and $f_+$. Setting
\bea
{1\over6}H_{abc}\Gamma^{abc}=\nu \Gamma^{+16}~,
\eea
the dilatino KSE can be written as
\bea
[\partial_+\Phi \Gamma^+-{1\over2} \nu  \Gamma^{+16}-{1\over2} f_+\Gamma^+ c(\omega)-{1\over2} c(S)]\e=0~.
\eea
If $\mu\not=0$, acting with $\Gamma^+$ on the above equation, one finds that
\bea
c(S)\Gamma^+\e=0
\eea
which in turn gives $\Gamma^+\e=0$. Substituting this into the KSE, one concludes that $c(S)\e=0$ and so for
 supersymmetric solutions $\mu=0$. Therefore the dilatino KSE becomes
 \bea
\partial_+\Phi \Gamma^+-{1\over2} \nu  \Gamma^{+16}-{1\over2} f_+\Gamma^+ c(\omega)]\e=0~.
\eea
Writing $\e=\e_-+\e_+$ with $\Gamma^+\e_+=0$, the solutions preserve at least 2 supersymmetries with Killing spinors
$\e_+$. To find whether more supersymmetries are preserved substitute $\e=\e_-+\e_+$ into the KSE and observe that
\bea
[\partial_+\Phi -{1\over2} \nu  \Gamma^{16}-{1\over2} f_+c(\omega)]\Gamma^+\e_-=0~.
\eea
Acting with $\Gamma^{16}$ on the above equation and taking eigenspaces with respect to $\Gamma^{16} c(\omega)$ and
observing that $(\Gamma^{16})^2=-{\bf 1}$, there are solutions provided that
\bea
\partial_a\Phi=0~,~~~\nu=\pm 3 f_+~.
\la{cwn4}
\eea
These are in fact the conditions for backgrounds with 4 supersymmetries.

\subsection{Geometry}

The geometry of the base space $B$ of the spacetime depends on whether $\mu$ and $f$ vanish.
If $\mu\not=0$, then the Nijenhuis tensor of the
base space does not vanish and so $B$ is an almost complex manifold. On the other hand, solutions with $\mu=0$
have as base
space a complex manifold.

The $SU(3)$ structure of the spacetime is {\it not} always inherited by the base space $B$.
If $f=0$, then both $\omega$ and $\chi$ are invariant under the infinitesimal
transformations generated by the Lie algebras (\ref{su3a}), ${\cal L}_a\omega={\cal L}_a\chi=0$.
 Since in addition they both vanish along the fibre directions, they
descent to a Hermitian form and a holomorphic (3,0)-form on the base $B$, respectively. Moreover, these data
are compatible with $(\tilde g, \tilde H)$. So $B$ is a manifold with an $SU(3)$ structure compatible
with a metric connection with skew-symmetric torsion $\hat{\tilde\nabla}$. Such geometries
have been investigated extensively in \cite{strominger}-\cite{chiossi}.

Next suppose that $f\not=0$. In this case again ${\cal L}_a\omega=0$ and it vanishes
along the fibre directions of the spacetime, thus it descents to a Hermitian form on $B$ and so $B$
is an almost complex manifold. (It becomes complex if $\mu=0$.) In addition $\omega$
is compatible with $\hat{\tilde\nabla}$, $\hat{\tilde\nabla}\omega=0$. This is not the case with $\chi$.
Although $\chi$ vanishes along the fibre directions of spacetime, the Lie derivative of $\chi$
does not (\ref{su3lc}). As a result $\chi$ does not descent as a (3,0)-form on $B$ but rather as a (3,0)-form
twisted by a line bundle. Moreover $\chi$ is not compatible with the data $(\tilde g, \tilde H)$ of the base space.
To see this,  write $x^M=(y^\alpha, x^\mu)$, where $x^\mu$ are  coordinates of the base space $B$
and $y^\alpha$ are coordinates of the fibre of $M$. Then
\bea
e^a\equiv \lambda^a=\lambda^a{}_\alpha dy^\alpha+ \lambda^a{}_\mu dx^\mu~,~~~e^i=e^i{}_\mu dx^\mu~.
\eea
The inverse frame is
\bea
\lambda_a=\lambda_a^\alpha \partial_\alpha~,~~~e_i=e_i^\mu \partial_\mu- \lambda^b_\nu e^\nu_i
\lambda^\alpha_b \partial_\alpha~.
\eea
In particular
\bea
\partial_\mu= e_\mu^i \partial_i+\lambda^a_\mu \partial_a~.
\eea
Thus
\bea
\hat\Omega_\mu{}^i{}_j=e^k{}_\mu \hat\Omega_{k,}{}^i{}_j +\lambda^a{}_\mu \hat\Omega_{a,}{}^i{}_j=
e^k{}_\mu \hat\Omega_{k,}{}^i{}_j+\lambda^a{}_\mu
H_{aj}{}^i~.
\eea
Then $\hat\nabla_\mu\chi=0$ implies that
\bea
\hat{\tilde \nabla}_\mu\chi= 3i \lambda^a_\mu f_a \chi~.
\eea
Thus $B$ does not inherit the $SU(3)$ structure of the spacetime but rather a $U(3)$ structure
if $\lambda^a_\mu f_a\not=0$.

Furthermore, the twisting of the fibre directions of the spacetime over the base space depend on $f$. If $f=0$,
${\cal F}$ takes values, as a 2-form, in $\mathfrak{su}(3)$ with gauge group one of the groups in (\ref{su3a}), ie it is
a Donaldson connection.
However if $f\not=0$,
then ${\cal F}$ takes values in $\mathfrak{u}(3)$  with again gauge group one of the groups in (\ref{su3a}), ie
it is a Hermitian-Einstein connection.
Some of the
geometric properties of these solutions are tabulated in table 2.

\begin{table}[ht]
 \begin{center}
\begin{tabular}{|c|c|c|c|c|}\hline
$\mathfrak{Lie}\,G/N$& $1$ &$2$&$3$&$4$
 \\
\hline\hline
 $\bR^{3,1}$ & ${\rm AC}, U(3)$ & ${\rm (A)C}, (S)U(3)$ & $-$& ${\rm C}, SU(3)$
 \\
 \hline
 $\bR\oplus\mathfrak{su}(2) $&$-$&$-$&$-$& ${\rm C}, U(3)$
\\
\hline
$\mathfrak{sl}(2,\bR)\oplus \bR$&${\rm AC}, U(3)$&${\rm AC}, SU(3)$ &$-$& ${\rm C}, U(3)$\\
\hline
$\mathfrak{cw}_4$&$-$&${\rm C}, (S)U(3)$&$-$& ${\rm C}, U(3)$
\\
\hline

\end{tabular}
\end{center}
\label{tab1}
\caption
{ \small $N$ is the number of supersymmetries and $\mathfrak{Lie}\,G$ is the Lie algebra
of isometries of the solutions.   The entries give information about the geometry of base space.
${\rm AC}$ stands for almost complex   and ${\rm C}$ for complex manifold, respectively.
 ${\rm (A)C}$ stands for either complex or almost complex. The groups $SU(3)$ and $U(3)$ denote
  the holonomy of $\hat{\tilde \nabla}$.  $(S)U(3)$ means that the holonomy of $\hat{\tilde \nabla}$
  is either contained in $SU(3)$ or $U(3)$. The entries $-$ do not occur.}
\end{table}

\subsubsection{$\bR^{3,1}$}

\vskip 0.2cm
\underline {$N=1$}
\vskip 0.2cm

We have seen that the parameters of solutions with $N=1$ supersymmetry are restricted  as in
(\ref{su3dd}), for $\Delta_1, \Delta_2\geq 0$
and it is required  that both $A$ and $B$ do not vanish.

One of the properties of the geometry of these backgrounds is that the base space $B$ of spacetime
is always {\it almost complex}. So see this first recall that if $\partial_a\Phi$ is null or zero, then the
solutions preserve at least 2 supersymmetries. Thus, we always have $(\partial_a\Phi)^2\not=0$.
Now if $\mu=0$, $\Delta_1^2$ is negative unless $\partial_a\Phi f^a=0$. Thus for solutions
to exist,  in addition $\partial_a\Phi f^a=0$. In such case $A=0$
and the solutions preserve at least 2 supersymmetries.

Another consequence of the analysis above is that the dilaton is linear for all $N=1$ backgrounds
along the fibre directions. This follows from the requirement that $(\partial_a\Phi)^2\not=0$.

It is also straightforward to observe that if $f=0$, then $B=0$ and so again the
 solutions preserve at least 2 supersymmetries.
{ \it Thus we have shown that for solutions with $N=1$ supersymmetry,  $B$  is an almost complex manifold with
${\rm hol}(\hat{\tilde \nabla})\subseteq U(3)$}.

\vskip 0.2cm
\underline {$N=2$}
\vskip 0.2cm

Solutions with 2 supersymmetries may or may not admit a complex base space $B$. There are two
classes of solutions with complex base space ($\mu=0$) the following:

\begin{itemize}

\item $f$ and $\partial_a\Phi$ null and co-linear, $f_a=\ell \partial_a\Phi$, $\ell\geq 0$, $\partial_a\Phi\not=0$.

\item $f$ and $\partial_a\Phi$ spacelike and orthogonal, $\partial_a\Phi f^a=0$, $\partial_a\Phi\not=0$.

\end{itemize}

There are also solutions with 2 supersymmetries and almost complex base space. Again, there are two classes
\begin{itemize}

\item $\partial_a\Phi$ spacelike and $f_a=\ell \partial_a\Phi$, $\ell\geq0$.

\item $\partial_a\Phi=0$, $f\not=0$.

\end{itemize}

Of course in all cases, apart from the last one, the parameters are required to satisfy (\ref{su3dd}).

Therefore the base space of backgrounds with 2 supersymmetries may or may not be a complex manifold and
 may or may not have
${\rm hol}(\hat{\tilde \nabla})\subseteq SU(3)$. So there is a large range of possibilities.
 Moreover, there are backgrounds
for which the dilaton is constant along the fibre directions but for these $B$ is almost complex and
 ${\rm hol}(\hat{\tilde \nabla})\subseteq U(3)$.

There are no backgrounds with 3 supersymmetries. For backgrounds with 4 supersymmetries, the dilaton is constant
along the fibre directions, $B$ is complex, and
 ${\rm hol}(\hat{\tilde \nabla})\subseteq SU(3)$.

\subsubsection{$\bR\oplus \mathfrak{su}(2)$}

All solutions in this case preserve 4 supersymmetries.  It follows from \cite{het1} that the base space $B$ is
complex\footnote{The conditions for solutions with 4 supersymmetries are that $B$ is complex and
$\partial_a\Phi=0$, $\theta_\omega=2\tilde d\Phi$, $H_{aij}=H_{akl} I^k{}_i I^l{}_j$ and
${1\over 3} \epsilon_a{}^{bcd} H_{bcd}-\omega^{ij} H_{aij}=0$.}
  but
 ${\rm hol}(\hat{\tilde \nabla})\subseteq U(3)$.

\subsubsection{$\mathfrak{sl}(2)\oplus \bR$}

\vskip 0.2cm
\underline {$N=1$}
\vskip 0.2cm

We have seen that the parameters of solutions with $N=1$ supersymmetry are restricted  as in
(\ref{n1su3}), for $2\mu\bar\mu\geq (\partial_6\Phi)^2$
and it is required  that both $A$ and $B$ do not vanish.

It is clear from the conditions of the dilatino KSE that for all $N=1$ backgrounds the base space is {\it
an almost complex manifold and ${\rm hol}(\hat{\tilde \nabla})\subseteq U(3)$}. However unlike the $\bR^{3,1}$ case,
there are backgrounds with 1 supersymmetry and constant  dilaton in
the fibre directions, $\partial_a\Phi=0$.

\vskip 0.2cm
\underline {$N=2$}
\vskip 0.2cm

The base space $B$ of solutions with 2 supersymmetries is always an almost complex manifold. However in this
case $f=0$. As a result the base space $B$  is {\it an almost complex manifold with
${\rm hol}(\hat{\tilde \nabla})\subseteq SU(3)$}. There are also solutions with constant dilaton along the
fibre directions $\partial_a\Phi=0$.

 For solutions with 4 supersymmetries $B$ is a complex manifold,
$f\not=0$. The
geometry has been investigated in \cite{het1}.

\subsubsection{$\mathfrak{cw}_4$}

There are backgrounds with 2 or 4 supersymmetries. In particular, there are no $N=1$ solutions. For all
$N=2$ solutions the base space is complex $\mu=0$.  Moreover, there are solutions with constant or non-constant
dilaton and with either $f=0$ or $f\not=0$, ie either ${\rm hol}(\hat{\tilde \nabla})\subseteq SU(3)$
or ${\rm hol}(\hat{\tilde \nabla})\subseteq U(3)$.

The base space of solutions with $N=4$ supersymetries is a complex manifold and ${\rm hol}(\hat{\tilde \nabla})\subseteq U(3)$.
This is because $f$ does not vanish as it can be seen in (\ref{cwn4}).

In all cases $SU(3)$ we have investigated above, there are no solutions with 3 supersymmetries. This is reminiscent to the absence of solutions
preserving 31 supersymmetries in IIB, IIA and 11-dimensional supergravities \cite{iib31, iia31}. However we have not ruled
out the possibility that solutions with 3 supersymmetries exist after an appropriate discrete identification
of solutions that preserve 4 supersymmetries as in \cite{gutowski}.

\newsection{ $SU(2)$}

\subsection{Holonomy reduction}

The backgrounds with ${\rm hol}(\hat\nabla)\subseteq SU(2)$ admit six 1-forms $e^a$ constructed from
parallel spinor bi-linears.  So the typical fibre of $TM$ decomposes as $\bR^{9,1}=\bR^{5,1}\oplus \bR^4$ and
$SU(2)$ acts on $\bR^4\otimes \bC=\bC^2\oplus \bar\bC^2$ with the fundamental representation and its complex conjugate.
Moreover a vector in $\bC^2$ (or $\bar\bC^2$) has isotropy group $\{1\}$ in $SU(2)$.  As in the previous
cases, the Bianchi identity (\ref{bbb}) and $dH=0$
imply that $[e_a, a_b]$ are $\hat\nabla$-parallel. Thus if one
of the vector field $[e_a, e_b]$ is linearly independent from those in $\{e_a\}$, then the holonomy
of $\hat\nabla$ reduces to $\{1\}$ and it becomes a special case of backgrounds with holonomy $\{1\}$. These
have been classified in \cite{josek}, see also \cite{het2}.  Thus to investigate $SU(2)$ backgrounds which
do not reduce to $\{1\}$ ones,  we shall take that the vector space spanned by $\{e_a\}$ to close
under Lie brackets. So  $\{e_a\}$'s span a 6-dimensional Lorentzian Lie algebra and these are isomorphic to
\bea
\bR^{5,1}~,~~~\bR^{3,1}\oplus \mathfrak{su}(2)~,~~~\mathfrak{sl}(2,\bR)\oplus \bR^4~,~~~\mathfrak{sl}(2,\bR)\oplus \mathfrak{su}(2)~,~~~
\mathfrak{cw}_4\oplus \bR^2~,~~~\mathfrak{cw}_6~.
\la{lor}
\eea

The fundamental forms of $SU(2)$ are the three Hermitian forms
\bea
\omega_1=- e^3\wedge e^8-e^4\wedge e^9~,~~~\omega_2=e^3\wedge e^4-e^8\wedge e^9~,~~~\omega_3=-e^4\wedge e^8+e^3\wedge e^9
\la{su2par}
\eea
therefore $a=0,5,1,6,2,7$. These are associated with endomorphism
which satisfy the algebra of imaginary unit quaternions, $I_r I_s=-\delta_{rs} 1_{4\times4}+\epsilon_{rst} I_t$.

To
identify the components of $\tilde d e^a$ along $\mathfrak{su}(2)^\perp$,  we use the decomposition
$\Lambda^2(\bR^4)=\mathfrak{su}(2)\oplus \mathfrak{su}^\perp(2)$, where $\mathfrak{su}^\perp(2)=\mathfrak{su}(2)$.
The $\mathfrak{su}(2)$ component is spanned by the anti-self-dual 2-forms while the $\mathfrak{su}^\perp(2)$
is spanned the self-dual 2-forms given in (\ref{su2par}). Therefore, the $\mathfrak{su}^\perp(2)$ component of
$\tilde d e^a$ can be written as
\bea
\tilde de^a+\star \tilde d e^a=2 f^a_r \omega^r~.
\eea
Since the Bianchi identity (\ref{bbb}) and $dH=0$ imply that $\tilde de^a+\star \tilde d e^a$ are $\hat\nabla$-parallel,
 $f^a_r$ are some real constants.

It can be easily seen that there is no a $\mathfrak{su}(2)$-invariant component of $\tilde H$, and so $S=0$.
Thus for the $SU(2)$ case, equation (\ref{lin}) is written as
\bea
\partial_a\Phi={\rm const}~,~~~[e_a, e_b]=-H_{ab}{}^c e_c~,~~~\tilde de^a+\star \tilde d e^a=2 f^a_r \omega^r~,
~~~S=0.
\la{linsu2}
\eea

As in the $SU(3)$ case, we have that $\partial_a\Phi$ and $f$ are restricted. In particular one finds that
\bea
\partial_a\Phi H^a{}_{bc}=0~,
\la{ph}
\eea
and
\bea
 -H^c{}_{ab} f_{cs}=2 f_{ar}f_{bt} \epsilon^{rt}{}_s~.
\la{hom}
\eea
The former equation follows from the field equation of the 2-form gauge potential as in the $G_2$ and $SU(3)$ cases.
To prove (\ref{hom})
use $\hat \nabla \omega^r=0$, the quaternionic algebra of $I_r$'s and $\tilde de^a+\star \tilde d e^a=2 f^a_r \omega^r$
in (\ref{linsu2}) to find  that
\bea
{\cal L}_a \omega^r= 2 f_{as} \epsilon^s{}_{rt}\, \omega^t~.
\eea
Moreover the property $[{\cal L}_X,{\cal L}_Y]={\cal L}_{[X,Y]}$ of the Lie derivative implies
 (\ref{hom}).
Therefore $f:~~\mathfrak{Lie}\, G\rightarrow \mathfrak{su}(2)$ is a Lie algebra homomorphism.

\subsubsection{Solution of the homomorphism condition (\ref{hom})}

Depending on $\mathfrak{Lie}\, G$ in (\ref{lor}), (\ref{hom}) has two non-vanishing solutions. One solution is
\bea
f_{ar}= w_a v_r~,~~~ v_r v^r=1~,
\la{wv}
\eea
provided that
\bea
H^c{}_{ab}\, w_c=0~.
\la{orth}
\eea
Clearly the direction along $w$ in $\mathfrak{Lie}\,G$ is central. So this solution exists for all $\mathfrak{Lie}\, G$
in (\ref{lor}) apart from $\mathfrak{sl}(2)\oplus\mathfrak{su}(2)$.

The other solution is
\bea
(f_{as})=(0,0,0, -{1\over2} H_{627} \tilde f_{r's})~,~~~r'=6,2,7~,~~~s=1,2,3~,
\la{iso}
\eea
provided that
  $\mathfrak{Lie}\, G=\mathfrak{h}\oplus \mathfrak{su}(2)$, $\mathfrak{su}(2)$ spans the directions $6,2,7$ and
  $(\tilde f_{r's})={\rm diag}(1,1,1)$. Therefore $f$ is  a Lie algebra
homomorphism with Kernel $\mathfrak{h}$. Thus  when it is restricted on the $\mathfrak{su}(2)$ subalgebra
of $\mathfrak{Lie}\, G$, it is
 a Lie algebra
isomorphism.

To prove (\ref{wv}) and (\ref{iso}),  first observe that if there is a $t\in \mathfrak{Lie}\, G$
which is central and $t^a f_{ar}\not=0$, then all the other $f_{ar}$
are proportional to $t^a f_{ar}$, ie $f_{br}= u_b\, t^a f_{ar}$.
This follows easily by contracting (\ref{hom}) with $t$.
In particular one has  $|v| v_r=t^af_{ar}$ in (\ref{wv}).

Moreover  if $\mathfrak{Lie}\, G$ contains a $\mathfrak{cw}$ algebra, all solutions  of (\ref{hom}) are of the
type (\ref{wv}). To show this,  observe that the non-vanishing
structure constants of $\mathfrak{cw}$ are of the type $H_{+ij}$. If $(f_{-r})$ does not vanish, then it follows from the
previous statement that the solution is (\ref{wv}). Next suppose that $f_{-r}=0$.
Setting $a=i$ and $b=j$ in (\ref{hom}) and using the only $H_{+ij}$ are non-vanishing, one finds that
$f_i=(f_{ir})$ are proportional to each other, ie $f_{ir}= u_i v_r$. Then setting $a=+$ and $b=j$ in (\ref{hom})
and using the proportionality of $f_i$'s,
it is easy to see that $f_+=(f_{+r})$ is also proportional to $f_i$'s. Thus the only solution is (\ref{wv}).

It remains to investigate (\ref{hom}) for the Lie algebras $\mathfrak{Lie}\, G=\mathfrak{h}\oplus \mathfrak{su}(2)$
in (\ref{lor}).
Now if  $t^a f_{ar}=0$ for all
$t$ central elements, then one derives (\ref{iso}) for the $\bR^{2,1}\oplus \mathfrak{su}(2)$ case.
Next consider
$\mathfrak{sl}(2,\bR)\oplus \mathfrak{su}(2)$.
Using the fact that $\mathfrak{sl}(2,\bR)$ and $\mathfrak{su}(2)$ commute, if $f_a=(f_{ar})$
has non-vanishing components for directions in both these subalgebras, then one concludes
the only solution is (\ref{wv}). But since $w$ is required in addition  to be central, this solution is excluded.
Alternatively $f_a=(f_{ar})$ must vanish when restricted on either $\mathfrak{sl}(2,\bR)$ or $\mathfrak{su}(2)$.
 In addition the Kernel must be $\mathfrak{sl}(2,\bR)$ since $f$ is a Lie
algebra homomorphism. Thus  $(f_{as})=(0, f_{rs})$, where $(f_{rs})$ is an invertible matrix.

To proceed, we can always choose $e^a$ such that $\eta_{ab}$ is the Minkowski metric. Without loss of generality,
we orient $\mathfrak{su}(2)$ in the directions $6,2$ and $7$, and so the structure constants
 are $H_{r's't'}=H_{627}\epsilon_{r's't'}$, $r',s',t'=6,2,7$, $\epsilon_{627}=1$. Next setting
\bea
f_{r'r}=-{1\over2} H_{627} \tilde f_{r'r}~,
\eea
one finds that
\bea
\tilde f_{r'r} \tilde f_{s's} \epsilon^{rs}{}_t=\epsilon_{r's'}{}^{t'} \tilde f_{t't}~.
\eea
It is easy to see that the 3-vectors $\tilde f_{r'}=(\tilde f_{r'r})$ have unit length and are mutually orthogonal.
Therefore
up to an orthogonal transformation, one can set $\tilde f_{61}=\tilde f_{22}=\tilde f_{73}=1$ and the rest of the components to vanish.
This  proves (\ref{iso}).

\subsubsection{Dilatino KSE}

Adapting the general analysis in section 2 to this case, the part of the dilatino
KSE involving the Lee form (\ref{leee}) is
\bea
 \theta_{\omega_1}=\theta_{\omega_2}= \theta_{\omega_3}=2\tilde d\Phi~.
 \la{su2lee}
\eea
Using (\ref{linsu2}), the rest of the KSE (\ref{sd}) can be  written as
\bea
(\Gamma^a\partial_a\Phi-{1\over12} H_{abc}\Gamma^{abc}-{1\over 4} f_{ar}\, \omega^r_{ij}\, \Gamma^{aij})\epsilon=0~.
\la{finsu2}
\eea
In what follows we shall solve (\ref{finsu2}) for the Lie algebras (\ref{lor}), and for the solutions of
the homomorphism
condition (\ref{wv}) and (\ref{iso}).

\subsection{$f=wv$}

The dilatino KSE in the (\ref{wv}) becomes
\bea
(\Gamma^a\partial_a\Phi-{1\over12} H_{abc}\Gamma^{abc}-{1\over 4}w_a v_r (\omega^r)_{ij} \Gamma^{aij})\epsilon=0~.
\la{finsu2a}
\eea
The solutions depend on the properties of $w$ and $\partial\Phi$, and so we shall examine various cases. We shall
always assume that $w\not=0$ since the $w=0$ case is examined in appendix A. We shall demonstrate that
there are solutions that preserve either 4 or 8 supersymmetries.

\subsubsection{$\bR^{5,1}$}

\vskip 0.2cm
\underline{$w^2\not=0$ }
\vskip 0.2cm

Since $\bR^{5,1}$ is abelian $H_{abc}=0$. Act on (\ref{finsu2a}) with $w_a\Gamma^a$ and write the resulting
equation as
\bea
(\partial_a\Phi w^a+A-w^2 B)\e=0
\la{www}
\eea
where
\bea
A=w_a\partial_b\Phi \Gamma^{ab}~,~~~B={1\over4} v_r (\omega^r)_{ij} \Gamma^{ij}~.
\eea
Observe that
\bea
A^2=-\big[w^2 (\partial_a\Phi)^2-(w^a \partial_a\Phi)^2\big]1_{8\times8}~,~~~B^2=- 1_{8\times8}~,~~~AB=BA~.
\eea
Since $B$ has always imaginary eigenvalues act with $B$ on (\ref{www}) to find
\bea
(\partial_a\Phi w^a B+AB+w^2 )\e=0~.
\eea
Separating the equation in eigenvalues of $AB$, there are solutions iff
\bea
\partial_a\Phi\, w^a=0~,~~~w^2 (\partial_a\Phi)^2>0~,
\eea
and
\bea
\pm\sqrt{w^2 (\partial_a\Phi)^2}=w^2~.
\eea
The first condition is required so that the dependence on $B$ to vanish because otherwise the eigen-spaces are complex.
The
second condition is required for the eigenvalues of $AB$ to be real. The last condition is necessary for matching the
eigenvalues of $AB$ with $w^2$. All these conditions have solutions, iff
both $w$ and $\partial_a\Phi$ are spacelike\footnote{The case that both are timelike is excluded
by the orthogonality condition.}  and orthogonal. So $w^2=(\partial_a\Phi)^2$. Note that if $\partial_a\Phi$ vanishes, for
consistency $w$ is null and this case is investigated below.
The solutions preserve 4 supersymmetries.

\vskip 0.2cm
\underline{$ w^2=0$ }
\vskip 0.2cm

 Acting on (\ref{finsu2}) with $\Gamma^aw_a$ as before, one finds that
\bea
(\partial\Phi \cdot w+ A)\epsilon=0
\eea
Acting now on (\ref{finsu2}) with $\Gamma^a\partial_a\Phi$  and using the above equation, one concludes that
\bea
((\partial_a\Phi)^2-2 \partial\Phi \cdot w { B})\e=0
\eea
This has no solutions unless $\inphi^2=\partial\Phi \cdot w=0$. Thus both $\partial_a\Phi$ and $w$ must be null and co-linear.

The analysis for the case that $\inphi^2=0$ is similar to that for which $w^2=0$ leading to the same solution that
$\partial_a\Phi$ and $w$ must be null and co-linear.
The solutions preserve 4 supersymmetries. The Killing spinors satisfy the projection $w_a\Gamma^a\e=0$.

\subsubsection{$\bR^{2,1}\oplus\mathfrak{su}(2)$}

\vskip 0.2cm
\underline{$ w^2\not=0$ }
\vskip 0.2cm

Define $A$ and $B$ as before and moreover set
\bea
C=w_a \partial_b\Phi\Gamma^{ab}-{1\over12} w_{a_1} H_{a_2a_3a_4} \Gamma^{a_1a_2a_3a_4}~.
\eea
Observe that the two terms in $C$ anti-commute with each other and
\bea
C^2=\Delta^2 \,1_{8\times 8}
\eea
where
\bea
\Delta^2=-\big[ w^2 (\partial_a\Phi)^2- (w^a \partial_a\Phi)^2]+{1\over4} w^2 H^2~,~~~H^2={1\over 6} H_{abc} H^{abc}>0~.
\eea
Moreover we have that
\bea
BC=CB~.
\eea
Acting on (\ref{finsu2}) with $w_a\Gamma^a$, observe that it can be rewritten as
\bea
(w^a\partial_a\Phi +C-w^2 B)\e=0~.
\la{sususu}
\eea
Since the eigenvalues of $B$ are imaginary, acting with $B$ on (\ref{sususu}) and separating the equation in eigenspaces
of $BC$, there are solutions iff
\bea
\Delta^2<0~,~~~w^a\partial_a\Phi=0~,~~~\sqrt{-\Delta^2}\pm w^2=0~.
\eea
The first condition is required for $BC$ to have real eigenvalues. Since $\partial_a\Phi$ is orthogonal
to $w$, $\Delta^2$ simplifies as
\bea
\Delta^2=-w^2 (\partial_a\Phi)^2+{1\over4} w^2 H^2~.
\eea

Now suppose that $w^2>0$. In such case $\Delta^2<0$ requires that $(\partial_a\Phi)^2>{1\over 4}H^2>0$. Thus
there are solutions that preserve 4 supersymmetries provided that $w$ and $\partial_a\Phi$ are both spacelike
and orthogonal and
\bea
(\partial_a\Phi)^2-{1\over4} H^2=w^2~.
\la{su2phw}
\eea
Next if $w^2<0$, the condition $\Delta^2<0$ requires that $(\partial_a\Phi)^2<{1\over 4}H^2$. Again there
are solutions preserving 4 supersymmetries provided that $\partial_a\Phi$ is spacelike. There are no solutions
with both $w$ and $\partial_a\Phi$ timelike because of the orthogonality condition. Note also that if $\partial_a\Phi$
is null, there is no solution.

\vskip 0.2cm
\underline{$ w^2=0$ }
\vskip 0.2cm

Acting with $w_a\Gamma^a$ on (\ref{finsu2}) and using $w^2=0$, one finds that
\bea
[w\cdot \partial\Phi+ w_a \partial_b\Phi \Gamma^{ab}-{1\over12} w_{a} H_{bcd} \Gamma^{abcd}]\epsilon=0
\eea
On the other hand, acting with $\partial_a\Phi \Gamma^a-{1\over12} H_{abc}\Gamma^{abc}$ on (\ref{finsu2}) and using the above equation, one finds that
\bea
[\inphi^2-{1\over24} H_{abc} H^{abc}-{1\over2} w\cdot \partial\Phi B ]\epsilon=0~.
\eea
This implies that
\bea
\inphi^2={1\over24} H_{abc} H^{abc}>0~,~~~w\cdot \partial\Phi=0~.
\la{connoc}
\eea

Since $\partial_a\Phi$ is required to be spacelike, we act on (\ref{finsu2}) with $\partial_a\Phi\Gamma^a$ to find
\bea
[(\partial_a\Phi)^2-{1\over12}\partial_a\Phi H_{bcd} \Gamma^{abcd}+ w_a\partial_b\Phi \Gamma^{ab}B]\e=0~.
\eea
Next observe that the last two terms in the above equation anti-commute and so define
\bea
D={1\over12}\partial_a\Phi H_{bcd} \Gamma^{abcd}- w_a\partial_b\Phi \Gamma^{ab}B~.
\eea
In terms of $D$ the dilatino KSE (\ref{finsu2}) can be written as
\bea
[(\partial_a\Phi)^2-D]\e=0~.
\eea
Observe that
\bea
D^2={1\over4} (\partial_a\Phi)^2 H^2 1_{8\times 8}~.
\eea
Therefore the eigenvalues of $D$ are real and so there are solutions which preserve 4 supersymmetries
provided that (\ref{connoc})
is satisfied. In fact the Killing spinors satisfy the projection $w_a\Gamma^a\e=0$. The remaining
dilatino KSE is also satisfied because the $SU(2)$-invariant spinors are chiral from the
6-dimensional perspective.

\subsubsection{$\mathfrak{sl}(2,\bR)\oplus \bR^3$}

The analysis of the solutions
of the dilatino Killing spinor equation is similar to that of the previous case. So we define again $A,B,C$ and $D$.
The only difference is that we now have to take into account that $\partial_a\Phi$ and $w$ are spacelike and $H$ is
timelike.

\vskip 0.2cm
\underline{$w^2\not=0$ }
\vskip 0.2cm

A brief inspection of the previous case reveals that there is a solution of the Killing spinor equations
preserving $4$ supersymmetries provided that both $\partial_a\Phi$ and $w$ are spacelike and orthogonal, and
(\ref{su2phw}) is satisfied. Observe that in this case, there are solutions for which $\partial_a\Phi=0$, ie
the dilaton is constant along the fibre directions.

\subsubsection{$\mathfrak{cw}_4\oplus \bR^2$ and $\mathfrak{cw}_6$}

Suppose now that $\mathfrak{Lie} G=\mathfrak{cw}_6$ or $\mathfrak{cw}_4\oplus \bR^2$. In this case $H$ is null. Acting with
$H_{abc}\Gamma^{abc}$ on (\ref{finsu2}), one finds that
\bea
\big(-{1\over12} H_{abc} \partial_d\Phi\Gamma^{abcd} +{1\over48} H_{abc} w_d \Gamma^{abcd} v_r \omega^r_{ij} \Gamma^{ij}\big)\epsilon=0~.
\eea
Next acting on the same equation with $\partial_a\Phi \Gamma^a-{1\over4} w_a\Gamma^a v_r\omega^r_{ij} \Gamma^{ij}$ and using the above
equation, one finds that
\bea
\inphi^2=w^2~,~~~\partial\Phi\cdot w=0~.
\la{su2cw46}
\eea
So there are two possibilities. Either $\partial_a\Phi$ and $w$ are null and co-linear or both are
spacelike and orthogonal.
In the former case, the backgrounds preserve $4$ of supersymmetries.
The Killing spinors satisfy the projection $w_a\Gamma^a\e=0$. Now if both $w=\partial_a\Phi=0$ and the structure
constants of $\mathfrak{cw}_6$ are self-dual, then the solution preserves 8 supersymmetries.

The latter case is only available
for $\mathfrak{Lie} G=\mathfrak{cw}_4\oplus \bR^2$. To investigate it further act on the dilatino KSE
with $w_a\Gamma^a$ and observe that it can be written as $(C-w^2 B)\e=0$ or equivalently as
\bea
(CB+w^2)\e=0~.
\eea
Moreover $(CB)^2= w^2 (\partial_a\Phi)^2 1_{8\times 8}$. Since both $\partial_a\Phi$ and $w$ are spacelike $CB$
has real eigenvalues and there are solutions preserving 4 supersymmetries provided that (\ref{su2cw46}) is satisfied.

\subsection{ Solutions for $f$ in (\ref{iso})}

\subsubsection{$\mathfrak{sl}(2,\bR)\oplus \mathfrak{su}(2)$ }

In the $\mathfrak{sl}(2,\bR)\oplus \mathfrak{su}(2)$ case, $\partial_a\Phi=0$. The KSE becomes
\bea
[-{1\over12} H_{abc}\Gamma^{abc}-{1\over4} f_{ar} \Gamma^a\omega^r_{ij}\Gamma^{ij}]\e=0~.
\eea
Without loss of generality, let us take $\mathfrak{su}(2)$ to be along the directions $051$ and $\mathfrak{su}(2)$ along the directions $627$.
In such case, the  Killing spinor equation can be rewritten as
\bea
\big[-{1\over2} [H_{051}+H_{627}] \Gamma^{627} -{1\over4} f_{r'r} \Gamma^{r'} \omega^r_{ij} \Gamma^{ij}\big] \e=0~.
\eea
Using (\ref{iso})
and acting with $\Gamma^{627}$, one finds
\bea
\big[{1\over2 } [H_{051}+H_{627}] -{\nu\over8} \epsilon_{r's't'} \Gamma^{s't'}  \omega^{r'}_{ij} \Gamma^{ij}\big] \e=0~,~~~
\nu=-{1\over2} H_{627}~,
\eea
where $\epsilon_{627}=1$ and $\omega^6\equiv \omega^1$, $\omega^2\equiv \omega^2$ and $\omega^7\equiv \omega^3$.
Next one can show that on the space of $SU(2)$-invariant spinors
\bea
W^2=-2W+3~,~~~~W={1\over8} \epsilon_{r's't'} \Gamma^{s't'}  \omega^{r'}_{ij} \Gamma^{ij}~.
\la{asu2}
\eea
Therefore $W$ has eigenvalues $1$ and $-3$. Since $W$ is traceless,
$1$ has multiplicity $6$ while $-3$ has multiplicity $2$. Therefore
if
\bea
H_{051}+2H_{627}=0
\eea
the background preserves 6 supersymmetries, while if
\bea
H_{051}-2H_{627}=0
\eea
the background preserves 2 supersymmetries. In the special case that the structure constants of $\mathfrak{sl}(2,\bR)\oplus \mathfrak{su}(2)$
are self-dual and $f=0$, the solutions preserve all 8 supersymmetries.

\subsubsection{$\bR^{2,1}\oplus \mathfrak{su}(2)$}

Next let us consider $\bR^{2,1}\oplus \mathfrak{su}(2)$. In this case, the Killing spinor equation can be written as
\bea
[Z+{1\over2} H_{627} -\nu W]\epsilon=0~,~~~\nu=-{1\over2}H_{627}~,
\eea
where $Z=\Gamma^{627} \partial_a\Phi \Gamma^a$ and $W$ is given in (\ref{asu2}). Since $\partial_a\Phi$ is  in the directions of $\bR^{2,1}$,
observe that
\bea
Z^2=\inphi^2\, 1_{8\times8}~,~~~ZW=WZ~.
\eea
Suppose that $\partial_a\Phi$ is spacelike, then the backgrounds preserve  4 supersymmetries provided that
\bea
\pm \sqrt{(\partial_a\Phi)^2}=-H_{627}~.
\eea
For each choice of sign, three of the Killing spinors belong to the eigenspace of $W$ with eigenvalue 1 and one Killing spinor
belongs to the eigenspace of $W$ with eigenvalue $-3$. In addition, they satisfy an appropriate projection with respect to $Z$.
There are no solutions for either $\partial_a\Phi$ null or timelike. All these backgrounds have linear dilaton along the
spacelike fibre directions of $\bR^{2,1}$.

\subsection{Geometry}

To solve the KSE (\ref{finsu2}) as in the $SU(3)$ holonomy case, we write $x^M=(y^\alpha, x^\mu)$,
where $x^\mu$ are  coordinates of the base space $B$
and $y^\alpha$ are coordinates of the fibre of $M$. Then, we have
\bea
\lambda^a=\lambda^a{}_\alpha dy^\alpha+ \lambda^a{}_\mu dx^\mu~,~~~e^i=e^i{}_\mu dx^\mu~,~~~
\partial_\mu= e_\mu^i \partial_i+\lambda^a_\mu \partial_a~,
\eea
as in the holonomy $SU(3)$ case.
Both the spacetime metric $g$ and $H$ are independent from the fibre coordinates $y$. Next as in \cite{callan} set
\bea
d\tilde s^2=h\, d\mathring{s}^2~,~~~~\tilde H=-\mathring{\star} dh~,
\la{4data}
\eea
where all tensors, including the function $h$, depend only on the coordinates $x^\mu$,
and the Hodge star operation is taken with respect to the metric $d\mathring{s}^2$.
To find the geometry of the base space $B$, one has to determine  $d\mathring{s}^2$.

Substituting the above data (\ref{4data}) into (\ref{su2lee}), we get
\bea
h^{-1}\partial_i h=2\partial_i \Phi~.
\eea
Thus
\bea
\partial_\mu \log h=2 (\partial_\mu \Phi- \lambda_\mu^a \partial_a \Phi)=0~.
\la{xx2}
\eea
Assuming that $\partial_a\Phi\not=0$,  one finds as an integrability condition that
\bea
{\cal F}^a_{\mu\nu}\, \partial_a\Phi=0~,
\eea
ie the curvature of the principal bundle connection must vanish along the direction
$\partial_a\Phi$ in $\mathfrak{Lie}\,G$.
Integrating locally (\ref{xx2}), we get
\bea
2\Phi=\log\,h+ Y(y)~,
\eea
where $Y$ depends only on the fibre coordinates. Since $\partial_a\Phi$ is constant, $\Phi$  is at most linear
in the $y$ coordinates.

To find the geometry of the base space, note that
\bea
\hat\Omega_\mu{}^i{}_j=e^k{}_\mu \hat\Omega_{k,}{}^i{}_j +\lambda^a{}_\mu \hat\Omega_{a,}{}^i{}_j= \mathring{\Omega}_{\mu,}{}^i{}_j+\lambda^a{}_\mu
H_{aj}{}^i~,
\eea
where $\mathring{\Omega}$ is the frame connection of the metric  $d^2\mathring{s}$ given in (\ref{4data}).
Since $\hat\nabla\omega_r=0$, the covariant derivative  of ${\omega}^r$ along $B$ with respect
to the Levi-Civita connection $\mathring{\nabla}$ of the
metric $d\mathring{s}^2$ is
\bea
\mathring{\nabla}_\mu(\omega_r)_{ij}-2 \lambda^a_\mu f_{as} \epsilon^s{}_r{}^t (\omega_t)_{ij}=0~.
\la{qut}
\eea
Note that if  $\lambda^a_\mu f_{as}\not=0$, then ${\cal L}_a\omega_r\not=0$. So the 2-forms $\omega_r$
 on $M$  do not descent
as 2-forms on the base space $B$. Instead, they are 2-forms on $B$ with values on a bundle with connection $\lambda$.

It is known that if there is no restriction on the connection $\lambda$, all 4-manifolds satisfy (\ref{qut}).
However in our case, there is a restriction because the self-dual part of the curvature of $\lambda$ satisfies
\bea
{\cal F}^a+\star {\cal F}^a=2 f^a_r \omega^r
\la{selfcon}
\eea
and $f$ is a constant matrix.

Clearly if $f=0$, the base space $B$ is hyper-K\"ahler and
the connection $\lambda$ an anti-self-dual instanton on $B$ with gauge group $\mathfrak{ Lie}(G)$.  On the other
hand if the anti-self dual part of ${\cal F}$ vanishes, then $B$ is self-dual Weyl Einstein.
This is the  Quaternionic K\"ahler condition in 4 dimensions, see eg \cite{salamon}.

Returning to the general case, the integrability condition  of (\ref{qut}) is
\bea
-\mathring{R}_{\mu\nu,}{}^ \lambda{}_\rho \mathring{\omega}^r_{\lambda\sigma}+\mathring{R}_{\mu\nu,}{}^ \lambda{}_\sigma \mathring{\omega}^r_{\lambda\rho}=
2{\cal F}^a_{\mu\nu} f_{as} \epsilon^{sr}{}_t \mathring{\omega}^t_{\rho\sigma}~.
\eea
This condition implies that the self-dual part of the Weyl tensor vanishes thus $B$ is a anti-self-dual Weyl manifold.
In addition, one finds  that
\bea
\mathring{R}_{\mu\nu,}{}^{ \rho\sigma} \mathring{\omega}^r_{ \rho\sigma}&=&4 {\cal F}^r_{\mu\nu}~,
\cr
\mathring{R}_{\mu\nu}&=&-2\sum_r {\cal F}^a_{\mu \rho} f_{ar} \mathring{\omega}^r_{\sigma\nu} \gamma^{\rho\sigma}~,
\eea
where $d\mathring{s}^2=\gamma_{\mu\nu} dx^\mu dx^\nu$. Observe that if the anti-self-dual part of ${\cal F}$
vanishes, then $B$ is  Einstein with cosmological constant proportional to $f^2$.

So far we investigated the general case.  Now we shall adapt the above results
to the cases (\ref{wv}) and (\ref{iso}).

\subsubsection{$f=wv$}

\vskip 0.2cm
\underline {$N=4$}
\vskip 0.2cm
To begin define
$\omega:=v_r \omega^r$. Then using (\ref{qut}) observe that $\omega$  satisfies
\bea
\mathring{\nabla} \mathring{\omega}=0~.
\eea
Thus $B$ is K\"ahler. Moreover, the remaining two equations in (\ref{qut}) become
\bea
\mathring{\nabla}_\mu\mathring{\omega}^r_{ij} -2 \lambda_\mu v_s \epsilon^{sr}{}_t \mathring{\omega}^t_{ij}=0~,~~~\lambda:= w_a\lambda^a~.
\eea
These two conditions are automatically satisfied for all the K\"ahler manifolds, in fact the other two $\omega$'s are sections of $K\otimes L$,
where $K$ the canonical bundle on $B$ and $L$ is the line bundle with connection $\lambda$. The only
additional condition is  the analogue of (\ref{selfcon}) which implies that
\bea
{\cal F}+\star {\cal F}=2 w^2 \omega~,
\la{selfcon2}
\eea
where ${\cal F}$ is the curvature of $\lambda$. In general $B$ is not Einstein.
But if in addition  the anti-self-dual part of ${\cal F}$ vanishes, then $B$ is K\"ahler-Einstein with non-vanishing cosmological constant, and the Ricci
tensor is given by
\bea
\mathring{R}_{\mu\nu}=2 w^2 \gamma_{\mu\nu}~.
\eea
The sign of the curvature of $B$  depends on whether $w$ is spacelike or timelike. If $w$ is null, then $B$ is hyper-K\"ahler. The only negative
curved case occurs whenever $\mathfrak {Lie}\, G=\bR^{2,1}\oplus \mathfrak{su}(2)$.

Thus, if $f=w v$, $B$ is {\it K\"ahler}. In particular ${\rm hol}({\mathring\nabla})\subseteq U(2)$.
Moreover the curvature of the connection along the
central element of $\mathfrak {Lie}\,G$ spanned by $w$,  $\lambda=w_a \lambda^a$,
satisfies the {\it Hermitian Einstein condition with cosmological constant}, ie the self-dual part is given by (\ref{selfcon2}). The remaining
components of ${\cal F}^a$ are {\it anti-self-dual}.

There is a special case of solutions with 4 supersymmetries and $B$ {\it hyper-K\"ahler}. These arise for example
whenever $w=0$. Moreover in this case ${\cal F}$ is {\it anti-self-dual}. Such solutions exist for all Lie algebras
in (\ref{lor}) unless their parameters are restricted such that the solutions admit 8 supersymmetries. Typically
the dilaton is linear along the fibre directions of spacetime but there are also $N=4$ solutions with constant
dilaton.

\vskip 0.2cm
\underline {$N=8$}
\vskip 0.2cm
The
 $N=8$ solutions have been classified in \cite{halfgp}. $B$ is {\it hyper-K\"ahler} and
 ${\cal F}$ is anti-self-dual with gauge group Lie algebra given in (\ref{lor}). The dilaton
 is constant along the fibre directions of spacetime.

\begin{table}[ht]
 \begin{center}
\begin{tabular}{|c|c|c|c|c|c|c|c|c|}\hline
$\mathfrak{Lie}\,G/N$& $1$ &$2$&$3$&$4$&$5$&$6$&$7$&$8$
 \\
\hline\hline
 $\bR^{2,1}\oplus \mathfrak{su}(2)$ & $-$ & $-$ & $-$& ${\rm HK}, {\rm K}, {\rm ASW}$& $-$& $-$& $-$& $-$
 \\
 \hline
$\mathfrak{sl}(2,\bR)\oplus \bR^3$&$-$&$-$ &$-$& ${\rm K}$& $-$& $-$& $-$& $-$\\
\hline
$\mathfrak{cw}_4\oplus \bR^2$&$-$&$-$&$-$& ${\rm HK},{\rm K}$& $-$& $$& $-$& $-$
\\
\hline
$\mathfrak{cw}_6$&$-$&$-$&$-$& ${\rm HK},{\rm K}$& $-$& $-$& $-$& ${\rm HK}$
\\
\hline
$\bR^{5,1}$&$-$&$-$&$-$& ${\rm HK},{\rm K}$& $-$& $$& $-$& ${\rm HK}$
\\
\hline
$\mathfrak{sl}(2,\bR)\oplus\mathfrak{su}(2) $&$-$&${\rm ASW}$&$-$& $-$& $-$& ${\rm ASW}$& $-$& ${\rm HK}$
\\
\hline
\end{tabular}
\end{center}
\label{tab1}
\caption
{\small  $N$ is the number of supersymmetries.   ${\rm ASW}$ (Anti-self-dual Weyl),
${\rm K}$ (K\"ahler) and ${\rm HK}$ (hyper-K\"ahler) refers to  the geometry of the 4-dimensional base space $B$ of the associated
supersymmetric background. The entries $-$ do not occur.
}
\end{table}

\subsubsection{$f$ given in (\ref{iso})}

\vskip 0.2cm
\underline {$N=2,4,6$}
\vskip 0.2cm

The analysis is identical to that given in the beginning  of the section for the general case. The only difference
is that $f$ is now restricted to be non-vanishing only on the $\mathfrak{su}(2)$ subalgebra
of $\bR^{2,1}\oplus \mathfrak{su}(2)$ and $\mathfrak{sl}(2,\bR)\oplus \mathfrak{su}(2)$. As we have already remarked
the self-dual part of the Weyl tensor of $B$ vanishes.
 In general, ${\rm hol}({\mathring\nabla})\subseteq Spin(4)$. The components ${\cal F}^r$
of ${\cal F}^a$ along the  $\mathfrak{su}(2)$ subalgebra
of $\bR^{2,1}\oplus \mathfrak{su}(2)$ and $\mathfrak{sl}(2,\bR)\oplus \mathfrak{su}(2)$ satisfy a Hermitian-Einstein
type of condition (\ref{selfcon}). The remaining anti-self-dual components of ${\cal F}^a$  are not restricted.

\newsection{Conclusions}

We have simplified the solution of the KSEs of heterotic backgrounds for which  the connection with skew-torsion
$\hat\nabla$  has holonomy  contained
in a compact group. In particular,
we have shown that if $dH=0$ and the field equations are satisfied, then
 there are restrictions on the fractions of supersymmetry that can occur and these depend
on the Lie algebra of isometries that acts on the backgrounds. We have collected an outline of our results
 in table 4.
This should be compared with table 2 of \cite{het3} which has been composed without making any additional assumptions.
The absence of $N=3$ solutions in the $SU(3)$ case and $N=7$ solutions in the $SU(2)$  case is a local result and reminiscent
to those in \cite{iib31, iia31} for 10- and 11-dimensional supergravities. We have not
 ruled
out the possibility that such backgrounds could exist after a discrete identification of solutions with 4 and 8
supersymmetries, respectively, see \cite{gutowski}. Observe in addition that there are no solutions in the $SU(2)$
case with 3 and 5 supersymmetries.

\begin{table}[ht]
 \begin{center}
\begin{tabular}{|c|c|}\hline
   ${\rm hol}(\hat\nabla)$ &$N$
 \\ \hline \hline
  $Spin(7)\ltimes\bR^8$& 1 \\
\hline
$SU(4)\ltimes\bR^8$&$\nearrow$, 2
\\ \hline
$Sp(2)\ltimes\bR^8$&$\nearrow$, $\nearrow$, 3
\\ \hline
$\times^2SU(2)\ltimes\bR^8 $&$\nearrow$, $\nearrow$, $\nearrow$, 4
\\ \hline
$SU(2)\ltimes\bR^8$&$\nearrow$, $\nearrow$, $\nearrow$, $\nearrow$, 5
\\ \hline
$U(1)\ltimes\bR^8$&$\nearrow$, $\nearrow$, $\nearrow$, $\nearrow$, $\nearrow$, 6
\\ \hline
$\bR^8$&$\nearrow$, $\nearrow$, $\nearrow$, $\nearrow$, $\nearrow$, $\nearrow$, $-$, 8
\\ \hline \hline
$G_2$& 1, 2
\\ \hline
$SU(3)$&1, 2, $-$ , 4
\\ \hline
$SU(2)$&$-$, 2, $-$, 4, $-$, 6, $-$, 8
\\ \hline
$\{1\}$& 8, 10, 12, 14, 16
\\ \hline
\end{tabular}
\end{center}
\caption{\small
In the  columns are the holonomy  groups that arise from the solution of the gravitino KSE
and the number $N$  of supersymmetries, respectively. $\nearrow$ and $-$ denote the entries
 in table 2 of \cite{het3} that
are special cases of backgrounds for which all parallel spinors are Killing and those that do not occur, respectively.  }
\end{table}

We have also examined the geometry of the solutions. In all these cases, the spacetime is a principal bundle
with fibre group $G$ and base space $B$. We have determined all the groups $G$ that occur and identify
the geometry of $B$ in all cases. In the $G_2$ case, $B$ admits a $G_2$ structure compatible
with a connection with skew-symmetric torsion. In the $SU(3)$ case, $B$ admits either a $SU(3)$ or a $U(3)$
structure again compatible with a connection with skew-symmetric torsion. $B$ is either
a complex or an almost complex manifold.
In the $SU(2)$ case, $B$ can either
be K\"ahler, or hyper-K\"ahler or anti-self-dual Weyl manifold. The latter condition
includes the 4-dimensional quaternionic K\"ahler manifolds which in addition are required to be Einstein.
 We have also found the conditions on the curvature ${\cal F}$ of the
connection that twists $G$ over $B$ imposed by supersymmetry. These are typically
instanton-like conditions in dimension 7, 6 and 4.  Furthermore, we examined the properties of the dilaton
for these backgrounds. In particular, we found that some of them exhibit a linear dilaton along the fibre $G$
of spacetime.

There is a worldvolume interpretation for all these solutions. In particular, they can be thought as
gauged WZW models with group manifold $G$ over $B$. The action that it is gauged is the left action which
is anomalous. However since the gauge fields are composite, this anomaly can be canceled from a contribution
form the base space. The details are similar to those in \cite{halfgp} as applied to
backgrounds that preserve 8 supersymmetries.

The supersymmetric heterotic backgrounds exhibit a rich geometric structure. Many explicit solutions are known, see eg
\cite{callan, gibbons, zamaklar, kostas}.
There is also a classification of all solutions with 8 supersymmetries \cite{halfgp} and a large class
of solutions with 4 supersymmetries  can be understood \cite{4gp}.
Further progress can be made to construct new examples. The main question is to find base manifolds
$B$ of the spacetime that have the prescribed properties required by supersymmetry. In most cases, there
are no general methods for the construction of such manifolds. The development of such methods is a
problem for the future.

\vskip 0.5cm
{\bf Acknowledgments:}~I would like to thank Ulf Gran,  Jan Gutowski and Andrew Swann for helpful discussions. I am
 partially supported
by the EPSRC grant, EP/F069774/1 and the STFC rolling grant ST/G000/395/1.

\setcounter{section}{0}

\appendix{Dilatino KSE on group manifolds}

It is instructive solve the dilatino KSE in the case that depends only on the dilaton and the Lie
structure constants $H_{abc}$. This has already been done for the group manifold
models in \cite{josek}. Here we shall adapt this to the ${\rm hol}(\hat\nabla)\subseteq SU(2)$ case. The holonomy
${\rm hol}(\hat\nabla)\subseteq SU(3)$
is straightforward. If $f=0$ in the ${\rm hol}(\hat\nabla)\subseteq SU(2)$ case, then  ${\cal F}$ is anti-self-dual.
The dilatino Killing spinor equation becomes
\bea
(\Gamma^a\partial_a\Phi-{1\over12} H_{abc}\Gamma^{abc})\epsilon=0~.
\la{finsu21}
\eea
There are
three cases to consider depending on whether $\partial_a\Phi$ is (i) $\inphi^2\not=0$, ie space- or time-like\footnote{
From now on we shall use the notation $\inphi^2=\eta^{ab} \partial_a\Phi \partial_b\Phi$ and $\apphi=\sqrt {|(\partial_a\Phi)^2|}$.} , (ii)
$\inphi^2=0$
but $\partial_a\Phi\not=0$, ie null, or (iii) $\partial_a\Phi=0$.

\vskip 0.2cm
\underline{$\inphi^2\not=0$}
\vskip 0.2cm

Acting with $\Gamma^a\partial_a\Phi$ on (\ref{finsu21}) and using (\ref{ph}), one finds that
\bea
\big[(\partial_a\Phi)^2-{1\over12} \partial_{a_1}\Phi\, H_{a_2a_3a_4}\Gamma^{a_1a_2a_3a_4}\big]\e=0~.
\eea
Next set
\bea
A={1\over12} \partial_{a_1}\Phi\, H_{a_2a_3a_4}\Gamma^{a_1a_2a_3a_4}
\eea
and  using that $H_{abc}$ are the structure constants of a metric Lie algebra observe that
\bea
A^2={1\over24} (\partial_a\Phi)^2 H_{abc} H^{abc} 1_{8\times 8}
\eea
For $A$ to have real non-vanishing eigenvalues
\bea
(\partial_a\Phi)^2 H_{abc} H^{abc}>0
\la{poss}
\eea
so $\partial_a\Phi$ and $H_{abc}$ are either both spacelike or both time-like.

There are no solutions if $H$ is timelike. To see this observe that  $H_{abc}$ is timelike
only when ${\mathfrak Lie}(G)=\mathfrak{sl}(2,\bR)\oplus \mathfrak{h}$ in (\ref{lor})
for $\mathfrak{h}=\mathfrak{su}(2)$
or $\bR^3$. For $\mathfrak{h}=\mathfrak{su}(2)$, (\ref{poss})  is incompatible with (\ref{ph}). Moreover
for  $\mathfrak{h}=\bR^3$, (\ref{ph}) implies
that $\partial_a\Phi$ is spacelike and so (\ref{poss}) is not satisfied.

It remains to see whether there are solutions if $H$ is space-like. If $H$ is space-like
 (\ref{poss}) is satisfied provided that $\partial_a\Phi$ is  space-like as well. In such case
 there are solutions  preserving
4 supersymmetries provided
\bea
\inphi^2={1\over24} H_{abc} H^{abc}>0.
\la{pos}
\eea
Moreover (\ref{finsu21}) is equivalent to the projector
\bea
\big(1-{\partial_{a_1}\Phi\, H_{a_2a_3a_4}\Gamma^{a_1a_2a_3a_4}\over12 (\partial_a\Phi)^2} \,\big)\e=0~.
\eea

Next observe that $A^2=0$, $A\not=0$, iff $H_{abc}$ is nilpotent.  It is easy to see from this that
consistency requires that $\partial_a\Phi$ must also be nilpotent. Thus such backgrounds  are included in the solutions
which shall investigate below. Therefore, the only solutions
 that can be arranged to  satisfied  the conditions (\ref{poss}) and (\ref{pos}), and so
admit  4 supersymmetries, have isometry algebra
\bea
\bR^{3,1}\oplus \mathfrak{su}(2)~.
\eea

\vskip 0.2cm
\underline{$\partial_a\Phi$ null}
\vskip 0.2cm

If $\partial_a\Phi$ is null, then $A\e=0$. Now acting on (\ref{finsu21}) with $H_{abc}\Gamma^{abc}$ and using
$A\e=0$, one finds that
\bea
H_{abc} H^{abc}\e=0
\eea
and so for solutions to exist $H_{abc}$ must also be null.

Without loss of generality the only non-vanishing component of $\partial_a\Phi$ can be taken to
be $\partial_+\Phi$. In such case (\ref{ph}) implies that the non-vanishing components
of $H_{abc}$ are $H_{+a'b'}$. In such case, the dilatino KSE is equivalent to the light-cone projection $\Gamma^+\e=0$.
The only groups listed in (\ref{lor})
that admit null structure constants are
\bea
\bR^{5,1}~,~~~
\mathfrak{cw}_4\oplus \bR^2~,~~~\mathfrak{cw}_6~.
\la{lor2}
\eea
As in the previous case, these backgrounds preserve strictly 4 supersymmetries.
\vskip 0.2cm
\underline{$\partial_a\Phi=0$}
\vskip 0.2cm

If the dilaton is constant, it is easy to see that (\ref{finsu21}) has solutions provided $H_{abc}$ is null.
The Lie algebras
\bea
\bR^{5,1}~,~~~\mathfrak{cw}_4\oplus \bR^2~,~~~\mathfrak{cw}_6~,
\eea
in (\ref{lor}) have null structure constants.
It turns out that the backgrounds with isometry algebras
\bea
\mathfrak{cw}_4\oplus \bR^2~,~~~\mathfrak{cw}_6~,
\eea
 preserve strictly 4 supersymmetries
provided that the structure constants of $\mathfrak{cw}_6$ are {\it not} self-dual, while
\bea
\bR^{5,1}~,~~~\mathfrak{cw}_6~,
\eea
preserve all 8 supersymmetries provided that the structure constants of $\mathfrak{cw}_6$ are self-dual.

\end{document}